\documentclass[11pt]{article}
\pdfoutput=1
\usepackage[utf8]{inputenc}
\usepackage{multirow}
\usepackage{amsmath, amsfonts, amssymb}
\usepackage{array}
    \newcolumntype{C}[1]{>{\centering\arraybackslash}m{#1}}
\usepackage{comment}
\usepackage{graphicx}
\usepackage{pdflscape}
\usepackage{psfrag}
\usepackage{amsthm}
\usepackage{enumerate}
\usepackage{arydshln}
\usepackage{pifont}
    
\usepackage{soul}
\usepackage{slashed}
\usepackage{subcaption}
\usepackage{mathrsfs}
\usepackage{ytableau}
 \usepackage{a4wide}
  \usepackage{tikz}
  \usepackage{tikz-cd}
  \usetikzlibrary{shapes.geometric}
  \usepackage{tcolorbox}
  \definecolor{dark-gray}{gray}{0.20}
  \definecolor{gray}{gray}{0.30}
  \definecolor{light-gray}{gray}{0.80}
  \definecolor{dark-red}{rgb}{0.7,0,0}
  \definecolor{dark-green}{rgb}{0.1,0.4,0}
  \definecolor{dark-blue}{rgb}{0.3,0.3,0.7}
  \definecolor{light-blue}{rgb}{0.8,0.8,1}
      \definecolor{swamp}{RGB}{240, 199, 197}
      
  \usepackage{pifont}

\usepackage{newunicodechar} 
\usepackage{setspace}

\usepackage{ifthen}

\usepackage{longtable}

\newcommand{\be}{\begin{equation}}
\newcommand{\ee}{\end{equation}}
\newcommand{\eq}[1]{(\ref{#1})}

\def\be{\begin{equation}}
\def\ee{\end{equation}}
\def\bea{\begin{eqnarray}}
\def\eea{\end{eqnarray}}

\newcommand{\C}{\mathbb{C}}
\newcommand{\R}{\mathbb{R}}

\newcommand{\Z}{\mathbb{Z}}

\AtBeginDocument{

}

\numberwithin{equation}{section}

\usepackage{jheppub}

\usepackage{cleveref}

\hypersetup{
	colorlinks=true,
	linkcolor=dark-blue,
	citecolor=dark-red,
	urlcolor=dark-green,
	linktoc=page
}

\theoremstyle{definition}

\theoremstyle{remark}

\crefname{appendix}{Appendix}{Appendices}

\title{\centering What IIB looks IIA string:\\ String Cobordisms via Non-Compact CFTs}
\author{Edoardo Anastasi,}
\author{Miguel Montero,} 
\author{Angel M. Uranga}
\author{and Chuying Wang } 
\affiliation{Instituto de F\'{i}sica Te\'{o}rica IFT-UAM/CSIC,
C/ Nicol\'{a}s Cabrera 13-15, Campus de Cantoblanco, 28049 Madrid, Spain}

\emailAdd{edoardo.anastasi@ift.csic.es, miguel.montero@csic.es, angel.uranga@csic.es, chuying.wang@ift.csic.es}
\abstract{The Swampland Cobordism Conjecture predicts the existence of end-of-the-world branes for every consistent Quantum Gravity theory, and domain walls connecting the Landscape. A perturbative string worldsheet description of these objects is only expected to exist when certain worldsheet invariants are vanishing or coincide across the domain wall. In this paper, we observe that many of these worldsheet obstructions can be evaded by allowing non-compact string worldsheets as part of the bordism. Using these ideas, we  provide a worldsheet QFT (flowing to a critical CFT under RG flow) that connects the worldsheets of 0A and 0B string theory, as well as those of IIA and IIB string theories. The non-compact character of these interpolations means that the description of the actual domain walls develop strongly coupled regions described by a linear dilaton background, where the worldsheet description breaks down. As a result, the IIA/IIB domain wall requires strong coupling region, in agreement with effective field theory considerations. Our constructions are heavily inspired by supercritical string theory, but are logically independent from it. We also analyze the fate of IIA/IIB NS5 branes as they cross the domain wall between these two theories. 
}

\begin{document}
\emergencystretch 3em
\hypersetup{pageanchor=false}
\makeatletter
\let\old@fpheader\@fpheader
\preprint{IFT-26-016}

\makeatother

\maketitle

\hypersetup{
    pdftitle={},
    pdfauthor={},
    pdfsubject={}
}

\newcommand{\remove}[1]{\textcolor{red}{\sout{#1}}}

\newcommand{\red}[1][\text{(check)}]{\textcolor{red}{#1}}

% Notation commands
\newcommand{\D}[1][\gamma]{\mathbf{D}_{\bf #1}}
\newcommand{\bvec}[1][\gamma]{\vec{b}_{\bf #1}}
\newcommand{\RFM}[2][T]{%
  \ifthenelse{\equal{#1}{T}}%
    {\frac{T^{#2}}{\Gamma}}%
    {\frac{\mathbb{R}^{#2}}{\mathcal{B}}}%
}
\newcommand{\Tr}[2]{{\rm Tr_{\bf #1}}(#2)}
\newcommand{\TrB}[1]{\Tr{B}{#1}}
\newcommand{\TrF}[1]{\Tr{F}{#1}}
\newcommand{\Vcas}{V_{\text{Cas}}}

\section{Introduction}
One of the cornerstones of the Swampland Program is the absence of global symmetries in consistent quantum gravity theories. As a particular instance of this general idea, reference \cite{McNamara:2019rup} introduced the Cobordism Conjecture, which posits that any quantum gravity vacuum must admit a boundary -- a ``wall to nothing'', similar to the $E_8$ end-of-the-world brane of M-theory \cite{Horava:1995qa,Horava:1996ma} (see also \cite{Montero:2025ayi}), where spacetime ends. Although these boundaries are codimension-1 loci in spacetime, the Cobordism Conjecture also leads to the existence of higher-codimension branes, since a $D$-dimensional quantum gravity compactified on a $k$-dimensional manifold leads to a $(D-k)$ dimensional quantum gravity, and the cobordism defect has codimension $(k+1)$ from the viewpoint of the $D$-dimensional theory. 

Many of the resulting spacetime boundaries are non-supersymmetric but still admit interesting worldvolume dynamics \cite{Kaidi:2024cbx,Yonekura:2024spl,Fukuda:2024pvu,Heckman:2024trz}. In some cases, like \cite{Kaidi:2024cbx}, novel non-supersymmetric string worldsheets describing the near-horizon geometry of cobordism defects were found, thus providing concrete evidence for the conjecture. One of the more exciting examples, the IIA/IIB domain wall (see \cite{Distler:2009ri} for a prophetic early reference), was recently studied in \cite{Heckman:2025wqd} from the spacetime and probe brane points of view, but no worldsheet description was provided. Relatedly, the analysis of \cite{Heckman:2025wqd} was carried out at the topological level, with no actual equations of motion being solved. This means that any dynamical aspects of the cobordism, such as the value of the tension or the gravitational backreaction of the domain walls on spacetime (particularly critical for codimension-1 objects) have not been taken into account (see e.g. \cite{Buratti:2021yia,Buratti:2021fiv,Angius:2022aeq,Blumenhagen:2022mqw,Blumenhagen:2023abk,Delgado:2023uqk} for dynamical cobordisms from the effective field theory viewpoint in other contexts).

The absence of a worldsheet description for the IIA/IIB domain wall and related examples such as the R7 brane \cite{Dierigl:2022reg,Dierigl:2023jdp,Debray:2023yrs} is perhaps for a good reason: The IIA/IIB domain wall is effectively a boundary condition for the chiral IIB spectrum, and this cannot happen at weak coupling. Since the IIB gravitini/dilatini and self-dual 4-form all have a gravitational anomaly, each of them individually does not admit a symmetry-preserving boundary condition \cite{Thorngren:2020yht}. Any boundary condition must involve all of them simultaneously. Given a valid boundary condition, we can proliferate it in spacetime, turning it into a symmetric mass generation mechanism for the IIB chiral fields. Realizations of symmetric mass generation have been explicitly described only in lower dimensions (see e.g. \cite{Razamat:2020kyf,Tong:2021phe} for 2d and 4d QFT discussions, \cite{Wang:2022ucy} for a review, and \cite{Angius:2024pqk} for 4d and 6d setups with gravity). Hence it is remarkable that the Cobordism Conjecture predicts the existence of such a mechanism in chiral 10d string theories, even though it has not been explicitly described as of now. At any rate, since it must happen at strong coupling, it is natural to expect that the IIA/IIB domain wall lacks a perturbative string description.

This does not mean perturbative string theory is not up to the task. Even if the core of the IIA/IIB domain wall is at strong coupling, there is hope of a perturbative worldsheet description slightly further away from the core. Indeed, the worldsheet description of heterotic cobordism defects in \cite{Kaidi:2024cbx} is precisely of this kind; there is a dilaton gradient growing towards the core of the solution, which is strongly coupled. The worldsheets of \cite{Kaidi:2024cbx} provide answers to many questions about the bordism defect, including their tensions \cite{Fukuda:2024pvu}, in spite of the lack of supersymmetry. We would like to have the same for type II cobordisms.

The goal of this work is to push the study of non-supersymmetric cobordisms as far as one can using the perturbative string description, and understand these obstructions from a worldsheet point of view. We will find that some obstructions to the existence of cobordism domain wall have a worldsheet avatar in that the corresponding worldsheet CFT's cannot be connected by up-and-down RG flows in the sense of \cite{Gaiotto:2019asa}. These obstructions are topological, and are detected by invariants such as the elliptic genus and TMF (topological modular form) classes \cite{Witten:1986bf, Gukov:2018iiq,Gaiotto:2019asa,Tachikawa:2021mvw,Tachikawa:2021mby,Tachikawa:2023nne,Tachikawa:2025awi,Moore:2025tmt}.

However, all the obstructions that we find are absent if we allow for non-compact CFT's to appear in the RG flow. In some concrete cases, such as the domain wall between 0A/0B string theories, or the IIA/IIB domain wall, we can find \emph{fully explicit} worldsheet descriptions of cobordism domain walls between these theories. The 0A/0B example actually already existed in the literature: the constructions in \cite{Hellerman:2007fc} naturally include an explicit interpolation between 0A and 0B string theories using supercritical string theory. We expand and give details on this configuration, following the interpretation as a supercritical string background\footnote{We note in passing that, although the no-ghost theorem is not proven for supercritical strings, there is no obvious obstruction to doing so; the theories amount to a simple modification of the correlation functions of the $X^0$ field. At any rate, the issue should be clarified.}. This involves the introduction/removal of worldsheet degrees of freedom via spacetime closed tachyon condensation (see \cite{Hellerman:2006nx,Hellerman:2006hf,Hellerman:2006ff,Hellerman:2007fc,Hellerman:2007ym,Hellerman:2007zz,Hellerman:2008wp,Hellerman:2010dv,Berasaluce-Gonzalez:2013sna,Garcia-Etxebarria:2014txa,Garcia-Etxebarria:2015ota,Angius:2022mgh} for related work), and the interpolation we construct corresponds to a kink tachyon profile (in close analogy with other closed tachyon solitons \cite{Berasaluce-Gonzalez:2013sna,Garcia-Etxebarria:2014txa,Garcia-Etxebarria:2015ota}). However, we also emphasize that the configuration can also be interpreted as a construction fully within critical string theory, where the non-compact bordism provides a QFT that flows to a critical CFT in the deep IR, in direct analogy with other  interpolations involving ``moving up-and-down the RG flow'' \cite{Gaiotto:2019asa}. We then proceed to do the same for the IIA/IIB domain wall; due to subtleties related to the GSO projection, we are forced to use a different construction, based on symmetric mass generation for systems of 8 2d Majorana-Weyl fermions via 4-fermion interactions \cite{Fidkowski:2009dba,Ryu:2012he,Qi:2012gjs,Tong:2019bbk} (a $Spin(7)$-preserving deformation of the Gross-Neveu model \cite{Gross:1974jv}), suitably extended to (1,1) supersymmetry. To our knowledge, this (1,1) extension is novel in the literature. Along the way we study the behavior of D-branes across the domain wall, following \cite{Heckman:2025wqd}, and include an analysis of NS5 branes, which was not carried out in that reference.

Allowing for non-compact CFT's for worldsheet bordisms is akin to allowing non-compact manifolds in ordinary geometric bordisms: The resulting bordism group is trivial since any $k$-dimensional manifold can be decompactified to $\mathbb{R}^k$. This simplification comes at a price:  After following the RG flow to its endpoint, we find that both the 0A/0B and IIA/IIB domain walls are described by a lightlike linear dilaton background (so in particular, they are not static), leading to a strongly coupled singular core. There are diverse proposed microscopic resolutions of such singularities, such as moving onto a supercritical phase \cite{Hellerman:2006ff}, or as a lightlike Big Bang \cite{Craps:2005wd} (although this later has been argued only in the type IIA case, it is reasonable to expect an analogous mechanism for IIB), describing a boundary akin to a lightlike Bubble of Nothing (BoN). Hence, we will remain fairly agnostic about the resolution of the strongly coupled core of our domain walls. However, for the purposes of the discussion, we will regard the simplest possibility that it is described by two back to back lightlike BoNs for the two sides, discussing interesting issues that can appear for type IIB light-like linear dilaton backgrounds in the Conclusions.
In any case, assuming this singularity is indeed a boundary, we can repeat the construction in any valid string compactification, thereby obtaining a general construction of Cobordism end-of-the-world branes in lightlike linear dilaton backgrounds. Although our solutions have very different kinematics from the familiar objects close to the supersymmetric vacuum, their existence suggests that, at least topologically, all cobordism classes admit a (in general, strongly coupled) boundary. 

The paper is organized as follows: In Section \ref{sec:obstructions} we review general worldsheet obstructions to the existence of cobordisms with a perturbative string description, and how they can be evaded by non-compact CFT's. In Section \ref{sec:type0} we study in detail the 0A/0B domain wall. In section \ref{sec:type0-worldsheet-interpolation} we describe the worldsheet interpolation; in section \ref{supercritical-type0} we realize it dynamically in terms of a closed tachyon kink in 11d supercritical type 0 theory; in section \ref{sec:type0-more} we describe the appearance of a 1d Majorana fermion in the interface defect on the worldsheet, and its role in relating charged and uncharged D-branes across the 0A/0B domain wall; in section \ref{sec:quantum} we obtain the infrared description describing the domain wall and argue it degenerates into two back-to-back lightlike linear dilaton backgrounds; in section \ref{sec:heterotic} we make an aside to provide a similar worldsheet description for the heterotic GSO domain wall introduced in \cite{Heckman:2025wqd}. Section \ref{sec:typeII} discusses the IIA/IIB domain wall, following similar techniques. In section \ref{sec:smg} we show the interpolation can be achieved by introducing 8-plets of additional 2d fermions and removing them via symmetric mass generation; in section \ref{sec:susy-smg} we present the full worldsheet interpolation by promoting the mechanism to (1,1) supersymmetry; in section \ref{sec:typeii-wall} we realize it dynamically to describe a physical type IIA/IIB domain wall; in section \ref{sec:typeii-supercritical} we explore its interpretation in terms of supercritical type II theories; and in section \ref{sec:strong-typeii} we discuss the strong coupling region and its possible UV resolutions. Section \ref{sec:NS5s} studies the fate of NS5-brane probes across the 0A/0B and IIA/IIB domain walls. In section \ref{sec:ns5-general} we study general features of the system, based on topology, symmetries and anomalies, without assumptions on the microscopic structure of the wall, and raise some puzzles; in section \ref{sec:ns5-supercritical} we exploit our microscopic descriptions of the domain walls to solve the anomaly puzzles. Finally, Section \ref{sec:conclusions} presents our conclusions and describes to what extent our results can be taken as evidence for the Cobordism Conjecture. Several details are relegated to Appendices.

\smallskip

{\bf Note added:} While completing this work, we were informed by E. Torres about the upcoming work \cite{Torres:2026vxx}, on a different approach to the description of type IIA/IIB domain walls.  

\section{Obstructions to worldsheet descriptions of cobordism domain walls}
\label{sec:obstructions}

The Cobordism Conjecture introduced in \cite{McNamara:2019rup} is a version of the absence of global symmetries in Quantum Gravity \cite{Banks:1988yz,Banks:2010zn} applied to semiclassical global charges that can be constructed out of spacetime topology. When applied to top-form global symmetries, it leads to the conclusion that all vacua of quantum gravity must admit a boundary, similar to e.g. the Horava-Witten M9 boundary in M-theory \cite{Horava:1995qa,Horava:1996ma}. The practical use of the conjecture is as follows: One starts with  a theory $\mathcal{T}$ that is a low-energy approximation to the fundamental quantum gravity. Most commonly, $\mathcal{T}$ is taken to be the low-energy effective supergravity or effective field theory. One then considers equivalence classes of manifolds which are valid backgrounds for $\mathcal{T}$, where two backgrounds are regarded as equivalent if there is a domain wall describable within $\mathcal{T}$ connecting one to the other. The corresponding set of classes for objects of a given codimension $k$ forms an abelian group, the cobordism group $\Omega_k^{\mathcal{T}}$. In general, the groups $\Omega_k^{\mathcal{T}}$ are non-zero signalling global symmetries present in $\mathcal{T}$, but which should be absent in the full quantum gravity. The idea is to consider a series of ever-increasing better approximations to the full theory,
\begin{equation} \ldots\longrightarrow\,{\mathcal{T}}\,\longrightarrow{\mathcal{T}'}\, \longrightarrow\ldots\longrightarrow\, \text{QG},\end{equation}
leading to a sequence of ever more refined bordism groups, which become trivial in the full theory of quantum gravity 
\begin{equation} \ldots\longrightarrow\,\Omega_k^{\mathcal{T}}\,\longrightarrow\Omega_k^{\mathcal{T}'}\, \longrightarrow\ldots\longrightarrow\, \Omega_k^{\text{QG}}=0.\end{equation}

The fact that we expect the cobordism group $\Omega^{\text{QG}}$ to vanish in quantum gravity does not mean that the cobordism groups in intermediate steps $\Omega^{\mathcal{T}}$ are useless. 
On the contrary, a non-vanishing cobordism class in $\Omega^{\mathcal{T}}$ means that the bordism defect that kills that class in quantum gravity must go beyond the corresponding approximation, and include features not present in $\mathcal{T}$.  
Any complete understanding of the corresponding bordism will require a more complete description $\mathcal{T}'$ that reduces to $\mathcal{T}$ at low energies. Said differently, $\mathcal{T}$-bordism invariants are obstructions to the description of a boundary within $\mathcal{T}$. As an illustrative example of this phenomenon, consider M-theory compactified on $\mathbb{RP}^4$ with 1/2 unit of $G_4$ flux \cite{Witten:1995em}. This class is non-trivial in $\mathfrak{m}^c$ bordism \cite{Witten:2016cio,Freed:2019sco},  which is the relevant structure to the 11-dimensional supergravity approximation to M-theory, so the object that has $\mathbb{RP}^4$ as a boundary (the OM5 plane \cite{Witten:1995em}) must be a singular object from the point of view of supergravity. 

Although in this example $\mathcal{T}$ is an effective field theory, there are also examples where the bordism invariant $\mathcal{T}$ goes beyond supergravity, like in perturbative string theory. For instance, consider $E_8\times E_8$ heterotic string theory compactified on a four-sphere $\mathbb{S}^4$, with an $E_8\times E_8$ bundle with 
\begin{equation} \int_{\mathbb{S}^4} c_2(E^{(1)}_8)=-\int_{\mathbb{S}^4} c_2(E^{(2)}_8)=k\neq0,\end{equation}
where $c_2(E^{(1)}_8)$ is the second Chern class of the first $E_8$ factor. 
This configuration is describable in supergravity and satisfies the heterotic Bianchi identity, and so it is a valid background for heterotic string theory. The class can be described in the string worldsheet by a sigma model with target space $\mathbb{S}^4$ and appropriate $E_8$ bundle; although such a construction will not be a worldsheet CFT, it will flow to one under RG flow. Since spacetime equations of motion correspond to vanishing worldsheet $\beta$ functions, we may regard the sigma model as an off-shell description of the cobordism class, that correctly captures the topological features but does not solve the actual equations of motion. From this point of view, it is natural to identify a perturbative string theory bordism class with a set of worldsheet CFT's that can be connected via RG flows. Importantly, the heterotic worldsheet must have $(0,1)$ supersymmetry for the GSO projection to be defined, and this condition must also hold off-shell.

As explained in \cite{Tachikawa:2021mby,Tachikawa:2021mvw,Tachikawa:2025awi}, and exploited in several recent papers (\cite{Kaidi:2023tqo,Kaidi:2024cbx,Saxena:2024eil} among others) precisely for heterotic string worldsheets the equivalence classes of $(0,1)$ SCFT's modulo up-and-down RG flows do form a spectrum (i.e. they have the mathematical structure that one would expect of a cobordism group), and in fact, there is a conjecture \cite{stolz-teichner} that it can be identified with the ring of topological modular forms (\emph{tmf}). This conjecture is powerful because \emph{tmf} is known, and so this can be used to partition heterotic worldsheets into deformation classes and look for  interesting worldsheet CFT's. Precisely this approach was followed in \cite{Kaidi:2023tqo,Kaidi:2024cbx} to construct new non-supersymmetric branes in heterotic string theory. One of the objects constructed is precisely the boundary of the $\mathbb{S}^4$ with $E_8\times E_8$ bundle described above. 
 This background is detected by the bordism invariant of heterotic $E_8\times E_8$ supergravity
\begin{equation} \int_{X} [\, c_2(E^{(1)}_8) - c_2(E^{(2)}_8)\,]\, ,\label{dee}\end{equation}
which happens to survive in the full $E_8\times E_8$ perturbative heterotic string theory as a non-trivial \emph{tmf} class. The existence of this invariant means that the theory $\mathcal{T}$ (the perturbative heterotic string) cannot describe the boundary of this configuration. Indeed, as was found in \cite{Craps:2005wd}, the near-horizon description of the bordism defect has a linear dilaton gradient, that grows strong as one approaches the core, signalling the breakdown of perturbative heterotic string theory. 

In this particular case, we can get a qualitative understanding of what is going on even beyond string perturbation theory, thanks to peculiarities of the background. Since we are only looking at topological features and are not requiring to obey equations of motion, we can consider making the $\mathbb{S}^4$ very large while the instanton of the first $E_8$ factor shrinks to a point in the north pole of $\mathbb{S}^4$ and that of the second $E_8$ shrinks to a point at the south pole. Locally, each of the instantons only sees an $\mathbb{R}^4$, and the shrinking we are describing is merely the small instanton transition \cite{Witten:1995gx}, where the instanton is described as an NS5-brane, or an M5 brane in the Horava-Witten description of heterotic strings \cite{Kaidi:2024cbx}. Since the two instantons have opposite charges, we actually have an M5-$\overline{\text{M5}}$ pair, where each brane is at a different $E_8$ boundary. To kill the cobordism class, we just have to bring them together. 

The fact that we had to consider a strong-coupling description to describe this cobordism defect is no accident, and can be ascertained from the low-energy supergravity (see \cite{Bergshoeff:2006bs,Angius:2024pqk} for the supergravity approximation to this configuration). The low-energy modes of  the original configuration include chiral, anomalous fermions charged under the subgroup of $E_8$ that remains unbroken by the instanton, and whose anomalies are cancelled by a six-dimensional Green-Schwarz mechanism. A weakly coupled (that is, lagrangian) description of the defect would imply gapping these chiral fermions and self-dual field in a Lagrangian fashion -- but there are no such terms. The only possible way to gap a fermion at weak coupling is via a mass term, which is forbidden by the fact that each of the massless chiral fermions involved are anomalous on their own. Instead, only the whole, anomaly-free system of the chiral fermions + self-dual field can possibly be gapped, and since this cannot be done at weak coupling, a strong coupling description is required, as emphasized in \cite{Angius:2024pqk}. In this case, by separating the two instantons, the system is very approximately supersymmetric, and we know the strong coupling physics that emerges: Once the gauge instanton becomes pointlike, the system is described by the strongly coupled $E_8$ SCFT, which mediates a tensor transition between the fermions and a self-dual tensor multiplet. This is an example of a symmetric mass generation mechanism \cite{Tong:2019bbk,Tong:2021phe,Wang:2022ucy}. After this point is reached, the low-energy description contains a self-dual and anti-self-dual tensor fields, which admit a mass term (corresponding to M5-$\overline{\text{M5}}$ annihilation). 

The morale of this story is that one should not expect a perturbative worldsheet description for cobordism defects where the low-energy effective field theory degrees of freedom cannot be gapped out in a lagrangian way. The fact that quantum field theory breaks down due to the infinite tower of string oscillator modes is no help, since all interactions switch off in the $g_s\rightarrow 0$ limit. 

This is true both for heterotic and type II strings. For instance, since the chiral degrees of freedom of IIB supergravity cannot be gapped out via mass terms, we would expect the IIA/IIB domain wall of \cite{Heckman:2025wqd} not to admit a worldsheet description. Although the analog of \emph{tmf} for type II strings is not known, whatever the right notion of perturbative type II bordism is should be a kind of up-and-down RG flow preserving the worldsheet $(1,1)$ supersymmetry. The IIA and IIB worldsheets cannot be connected like this: For any $(1,1)$ SQFT, one can forget the right-moving supercharge and compute the elliptic genus
\begin{equation} \text{tr}((-1)^{F^{ws}_L}\, q^{L_0-\frac{c_L}{24}} \bar{q}^{\bar{L}_0-\frac{c_R}{24}}),\end{equation}
where the trace is taken in the left-moving Ramond sector, and where $(-1)^{F^{ws}_L}$ is the left-moving worldsheet fermion number. As explained e.g. in \cite{Kaidi:2019pzj}, the difference between IIA and IIB can be recast as a replacement $(-1)^{F^{ws}_L}\rightarrow -(-1)^{F^{ws}_L}$. Therefore, the elliptic genus flips sign between IIA and IIB. Since the elliptic genus is invariant under $(1,1)$ RG flows, the IIA/IIB domain wall is not describable in the worldsheet. The same argument holds for the 0A/0B domain wall, although there is no longer a spacetime expectation for strong coupling since the low-energy fields of both 0A and 0B are all bosonic. 

Finally, we comment on the relationship between the up-and-down RG flows and conformal interfaces, which is another standard way in which to CFT's may be connected. Given some symmetry $G$ preserved by an up-and-down RG flow can be used to construct a $G$-preserving conformal interface (see \cite{Gaiotto:2012np} for an example), though the converse has not been established in general (see \cite{Brunner:2015vva} for a discussion in the context of perturbation theory). In  some examples, a conformal interface can turn out to be topological (transparent to the stress-energy tensor), in which case it becomes a symmetry operator that connects the same Virasoro CFT to itself. An example are the duality defects of \cite{Angius:2024evd}. Although a topological defect connects two theories that are identical as Virasoro CFT's (i.e. have the same conformal primaries and dimensions), the defect may or may not preserve global symmetries of interest. For instance, although there is a duality defect flipping the sign of the GSO projection \cite{Chang:2022hud}, it does not come from an up-and-down RG flow preserving fermion number; equivalently, the action of the defect on fermion number is non-trivial. In string perturbation theory, the physics is sensitive to the precise sign of $(-1)^{F_L^{ws}}$, so this distinction is physically relevant.

\subsection{A cheat: Non-compact bordisms}
\label{sec:cheat}

We have just seen that, given a theory $\mathcal{T}$, $\mathcal{T}$-bordism invariants provide an obstruction to constructing boundaries within $\mathcal{T}$. As explained above, the usual way to overcome this to pass to a more complete theory $\mathcal{T}'$ that recovers $\mathcal{T}$ at the low-energy limit. There is, however, another way, that to our knowledge has not been discussed in the literature\footnote{But is often brought up in informal discussion.}: Keep $\mathcal{T}$, but allow for deformations that would cost infinite energy in the construction of the bordism. 

The simplest example of this is perhaps an instanton in non-abelian gauge theory on $\mathbb{R}^4$. The instanton number is a topological invariant, provided that one restricts to finite-energy deformations only. If this restriction is lifted, then the instanton can be unwound, since all vector bundles on a contractible space such as $\mathbb{R}^4$ are trivial. A more relevant example for the Cobordism Conjecture is given by e.g. Spin cobordism, which arises as a first approximation to many quantum gravity vacua. All Spin bordism groups can be trivialized if we allow for bordisms involving non-compact manifolds. For instance, consider a circle $\mathbb{S}^1$ with periodic spin structure, which is non-trivial in spin bordism. Consider decompactifying $\mathbb{S}^1$ to $\mathbb{R}$, which has a single spin structure, and can then be re-compactified to an $\mathbb{S}^1$ with antiperiodic spin structure, which is trivial in spin bordism, and then to a point (see a more precise discussion in Appendix \ref{app:spin-structures}). We can do the same with any $k$-dimensional Spin manifold: Decompactify to $\mathbb{R}^k$, and then recompactify. Spin bordism (and any generalization based on manifolds decorated with some structure) can be completely trivialized in this way.

There is a worldsheet analog for these decompactifications: In the intermediate steps up-and-down RG flows discussed before, allow for \emph{non-compact} CFT's. The connection between $(0,1)$ SQFT's and \emph{tmf} is reliant on compact intermediate CFT's, as explained in \cite{Tachikawa:2025awi}. If this requirement is dropped, worldsheet invariants like elliptic genera can change continuously, due to leaking of non-normalizable groundstates at infinity \cite{Troost:2017fpk}. As a result, they cease to provide topological obstructions. Very few worldsheet obstructions survive this procedure. One example is the gravitational anomaly (the difference $c_L-c_R$ between left-moving and right-moving central charges), which in spacetime terms corresponds to the dimension of the compactification manifold. This matches the fact that, in the decompactifications considered in the previous paragraph, the dimension of the manifolds involved did not change.

The main objection to this kind of procedure is that, much like in the instanton example, decompactifying manifolds or worldsheet CFT's in this manner intuitively involves infinite energy, and may result in a cobordism defect of infinite tension, in contradiction with the Cobordism Conjecture, which does require finite tension for a valid defect. However, it is hard to make this precise without solving the equations of motion in detail: there are examples of localized bordism defects of finite tension where the core of the defect seems to decompactify, like the T-dual of a bubble of nothing \cite{Delgado:2023uqk}.

In the rest of this paper, we will explore how far we can get in constructing a worldsheet description for the IIA/IIB or 0A/0B domain wall using non-compact CFT's; we will find a close connection to supercritical string backgrounds, which were our original source of inspiration. Ultimately, however, we will find that after following the RG flow the resulting domain walls lead to linear dilaton backgrounds, thus leaving the regime of perturbative string theory as originally expected. From the point of view of the worldsheet, a strongly coupled region corresponds to infinte tension, as one expects the energy of the core to diverge as $g_s\rightarrow0$. While we will speculate of what lies on the other side of these strongly coupled regions, in some cases we will not be able ascertain whether they are Cobordism Conjecture boundaries.

\section{The 0A/0B domain wall}
\label{sec:type0}

In this section we illustrate the above general ideas in a concrete example, namely the construction of a worldsheet cobordism between 10d type 0A and 0B string theories. We will construct a worldsheet interpolation between the two theories, at the price of passing through a non-compact CFT. This may be regarded as a type 0 cousin of the type IIA/IIB GSO domain wall considered in \cite{Heckman:2025wqd}, so we refer to it as GSO wall as well. The extension to a similar construction for a type IIA/IIB GSO wall will be discussed in section \ref{sec:typeII}.

\subsection{Worldsheet description}
\label{sec:type0-worldsheet-interpolation}

The 10d type 0A and 0B string theories are described by a $(1,1)$ worldsheet theory with (ignoring ghosts) 10 bosons $X^\mu$ parametrizing the spactime coordinates and 10 Majorana fermions, whose left- and right-moving components we denote by ${\tilde \psi}^\mu$, $\psi^\mu$, respectively.
There is a $\Z_2$ gauge symmetry given by $(-1)^{F_{ws}}$, where $F_{ws}$ is the overall (left plus right) worldsheet fermion number. 
Equivalently the partition function implies a sum over spin structures, which are shared by left- and right-fermions, resulting in a diagonal modular invariant partition function. 

The difference between the 0A and the 0B theories is the choice of the fermion number of e.g. the left-moving Ramond groundstates. In more formal terms, this difference amounts to the stacking of a 2d  topological quantum field theory (TQFT), corresponding to $(-1)^{\rm Arf}$, on the left-moving sector, resulting in a sign flip in the fermion number of its Ramond groundstate. The Arf invariant (see e.g. \cite{Karch:2019lnn} for further information) is a $\Z_2$-valued topological invariant, given by the number of fermion zero modes mod 2 of the 2d theory on an arbitrary Riemann surface. On the torus, the Arf invariant is 0 except for the doubly periodic spin structure, hence accounting for the 0A/0B GSO sign flip in the partition function familiar from the textbook construction of the two theories.

Let us now proceed to define a worldsheet cobordism between these two theories. It is well known (see e.g. \cite{Karch:2019lnn}) that the stacking of the Arf TQFT is closely related to the introduction of one extra 2d massive Majorana fermion. More precisely, the stacking or not of the Arf TQFT is determined by the sign of the mass of the fermion. This allows to provide a simple interpolation between 10d type 0A and 0B worldsheet theories, by extending them with an extra Majorana fermion, with left- and right-moving components denoted by ${\tilde \psi}$, $\psi$, and a 2d bosonic partner $X$ to fill out a $(1,1)$ supermultiplet (denoted also by $X$, with some abuse of language). We now introduce a $(1,1)$ quadratic superpotential
\bea
W\sim \mu X^2\, ,
\label{supo-lambda}
\eea 
(from now on, we use subindices for the spacetime coordinates, so that superindices indicate exponents). In components, this introduces a mass term $\sim\mu^2X^2$ for the bosons and $\sim\mu {\tilde\psi}\psi$ for the fermions. Hence the above coupling makes the extra degrees of freedom massive, so the theory for any $\mu\neq 0$ flows to a 10d type 0 theory. But the fermion mass term flips sign as we cross from $\mu<0$ to $\mu>0$, hence the deformation interpolates between the worldsheet theories of 10d type 0A and 0B strings. Even though the extra degree of freedom $X$ is gapped for generic $\mu$, the family interpolates between two topologically different theories because is crosses the point $\mu=0$ at which the gap closes, opening up the possibility of changes of topological phase of the 2d theory, in agreement with the general arguments in section \ref{sec:obstructions}. As explained in \cite{Ooguri:2024ofs}, the vanishing gap means that the theory is becoming non-compact.

Our above procedure therefore can be regarded as starting in e.g. 10d type 0A theory, moving up an RG flow by integrating in a massive multiplet, moving in parameter space to flip the sign of its mass term, and moving down the RG flow by integrating it out to reach 10d type 0B theory. In this description, the crossing of $\mu=0$ seems illegal from the usual perspective of interpolations among CFTs. However, in the following we will explicitly show a perfectly well defined physical realization of the above worldsheet interpolation, in which the motion in parameter space is physically realized as motion along a spacetime coordinate, leading to a worldsheet description of a spacetime cobordism domain wall between the 10d 0A and 0B theories.

\subsection{Physical realization of the 0A/0B domain wall, and 11d supercritical type 0}
\label{supercritical-type0}

The physical realization of the above worldsheet interpolation can be described in terms of closed tachyon condensation in supercritical type 0 theory \cite{Hellerman:2006ff,Hellerman:2007fc}, whose basic ingredients are reviewed in appendix \ref{app:type0}. This viewpoint is not essential, and one may simply maintain the viewpoint of 10d critical string theory with additional gappable degrees of freedom on the worldsheet, but for convenience of the presentation we use the supercritical perspective. The connection with a pure 10d viewpoint is recovered after integrating out the extra degrees of freedom, and  will be emphasized at various points, in particular in section \ref{sec:quantum}.

We focus on 11d supercritical type 0 theory, following a construction in \cite{Hellerman:2007fc} which includes precisely the necessary ingredients used in our above discussion (generalization to more supercritical dimensions is straightforward, and we skip its discussion). The basic intuition is that 11d supercritical type 0 theory provides a physical realization for the additional 2d multiplet $X$, which now describes a physical 11th coordinate and its 2d fermion partner (in addition, the theory must include a timelike linear dilaton to cancel the extra contribution to the central charge). 

As described in appendix \ref{app:type0}, it is possible to remove the extra supercritical coordinate by a closed tachyon condensation process, which can be studied in an $\alpha'$-exact way if the tachyon profile is lightlike. Specifically, it is possible to introduce an exactly marginal deformation given by a lightlike tachyon profile (\ref{tachyon-quench10}) 
\bea
{\cal T}=\mu \exp(\beta X^+) X^2\, ,
\label{tachyon-quench10}
\eea
where $\mu$ is an arbitrary coefficient, and $\beta$ is fixed so that the tachyon deformation is an exactly marginal operator of the CFT. This couples as a $(1,1)$ superpotential (\ref{wsupo}), hence for late $X^+$ the above deformation  describes a mass term for the 11th coordinate supermultiplet, precisely as in (\ref{supo-lambda}). Depending on the sign of the mass term $\mu$, the tachyon condensation produces either the 10d type 0A or 0B theory. Interestingly, quantum corrections from integrating out the massive 2d multiplets modify the background metric and dilaton, turning the background into a lightlike linear dilaton, consistently with the fact that the endpoint theory is critical. This will be relevant for our analysis in section \ref{sec:quantum}.

The fact that the 11d type 0 theory is able to produce either the 10d type 0A or 0B theories is because it naturally encompasses both. The NSNS spectrum of the 11d type 0 theory contains the 11d graviton, 2-form and dilaton, as well as the closed tachyon. On the other hand, there is no chirality in 11d, so the Ramond groundstates  for left- and right-moving sectors transform in the unique spinor representation ${\bf 16}$ of the $Spin(9)$ Lorentz group in light-cone gauge. Hence, there is a unique 11d type 0 theory (with no A and B variants). The RR sector contains states arising as bispinors ${\bf 16}\otimes{\bf 16}$, which decompose into $p$-form potentials of all degrees, including even and odd. In terms of their 10d decomposition, we have bispinors 
\bea 
{\bf 16}\otimes{\bf 16}\;\rightarrow \; ({\bf 8_S}\oplus{\bf 8_C})\otimes ({\bf 8_S}\oplus{\bf 8_C})=({\bf 8_S}\otimes{\bf 8_C}\,\oplus\, {\bf 8_C}\otimes{\bf 8_S})\, \oplus\, ({\bf 8_S}\otimes{\bf 8_S}\,\oplus\, {\bf 8_C}\otimes{\bf 8_C})\, .\nonumber
\eea 
Namely, the RR field content of {\em both} the 10d type 0A and 0B theories. The closed tachyon condensation process selects one of the two possibilities by removing half of the field content. The latter process is difficult to describe from the spacetime point of view, given the lack of a fully satisfactory spacetime effective action including the tachyon (see Appendix \ref{app:rrforms} for a proposal encoding the key topological ingredients of the endpoint configurations), but is explicitly quantitative from the worldsheet perspective. The choice of 10d type 0A or 0B as the endpoint of closed tachyon condensation is simply determined by the overall sign of the tachyon profile.

This suggests a straightforward physical spacetime realization of the worldsheet interpolation between the two 10d theories. It corresponds to a configuration in which the mass term induced by the tachyon condensate flips sign as we move in one of the 10d spacetime coordinates, say $X_9$. Since this corresponds to a flip of the tachyon sign, we refer to it as tachyon kink. The simplest tachyon kink is of the form
\bea
{\cal T}\sim X_9 X^2\, ,
\label{tachyon-kink-type0}
\eea 
where we have momentarily ignored the exponential lightlike dependence. Intuitively, for generic $X_9$ we have a mass superpotential for the 11th coordinate $X$, and the fermion mass changes sign as we move across $X_9=0$. Hence, this provides a worldsheet description of a spacetime cobordism wall between 10d type 0A and 0B theories, located at $X_9=0$. In general, $X_9$ could be replaced by any function of $X_9$ with a simple zero at $X_9=0$, and in particular the choice ${\cal T}\sim \arctan X_9 X^2$ makes the superpotential asymptote to (\ref{supo-lambda}) for negative or positive constant $\mu$ at $X_9\to \mp\infty$, describing a `kink' between `constant' tachyon condensations with opposite signs. The linear dependence 
(\ref{tachyon-kink-type0}) is the leading approximation near $X_9=0$ and yields the key physics near the domain wall, so we will focus on that case.

There is another reason to focus on the linear dependence  on $X_9$ in (\ref{tachyon-kink-type0}), which is that it corresponds to a marginal deformation of the worldsheet CFT at leading order. To see this, notice that the two-derivative equation of motion for the type 0 tachyon is \cite{Hellerman:2006ff}
\begin{equation}\partial^2\mathcal{T}-2 V^\mu\partial_\mu \mathcal{T}+\frac{2}{\alpha'}\mathcal{T}=0,\quad V^+=V^-=\frac{q}{\sqrt{2}},\end{equation}
where $q^2=(D-10)/(4\alpha')$, we are working in light-like coordinates $(X^-,X^+,X^1,\ldots X^8)$, and we have introduced a linear dilaton gradient $\Phi=V^\mu X_\mu$ which only has non-vanishing $V^+$   and $V^-$ components (see \cite{Hellerman:2006ff} for details; the point is that these ingredients allow us to create a one-loop exact background with the correct central charge). If a tachyon perturbation obeys this equation, that means it is a worldsheet marginal deformation to first order in $\alpha'$. Expanding in transverse coordinates, the equation of motion becomes
\begin{equation} \partial_i^2\mathcal{T}-2\partial_+\partial_-\mathcal{T} -\sqrt{2}\, q (\partial_+\mathcal{T}+\partial_-\mathcal{T})+\frac{2}{\alpha'}\mathcal{T}=0\, .\end{equation}
The solutions in  \cite{Hellerman:2006ff}, of which our solution is an example, describe light-like domain walls where $\partial_-\mathcal{T}=0$. In this case the equation simplifies to
  \begin{equation} \partial_i^2\mathcal{T}-\sqrt{2}\, q\partial_+\mathcal{T}+\frac{2}{\alpha'}\mathcal{T}=0,\label{rerre}\end{equation}
  which \cite{Hellerman:2006ff} solves by taking
  \begin{equation}\mathcal{T}= \exp(\beta X^+) f(X^+,X^i),\end{equation}
  which turns \eq{rerre} into
    \begin{equation} \partial_i^2f-\sqrt{2}\, q(\partial_+ f + \beta f)+\frac{2}{\alpha'}f=0.\label{rerre2}\end{equation}
With one single extra dimension $X$ to condense, a solution is
   \begin{equation} f= \mu( X^2 + \frac{\sqrt{2}}{q} X^++c), \quad q\beta\equiv \frac{\sqrt{2}}{\alpha'}.\end{equation}
 Classically, this is a solution localized at $X=0$, with an ever-growing tachyon vev. Just as in the bosonic string analysis of   \cite{Hellerman:2006ff}, this classical term is cancelled by a one-loop renormalization (the theory is one-loop exact) to change the dilaton gradient and reduce the background to a $(D-1)$-dimensional one, without a tachyon vev. More concretely, since the scalar potential is
  \begin{equation} V\sim G^{\mu\nu} \partial_\mu\mathcal{T}  \partial_\nu\mathcal{T} = G^{ij} \partial_i f  \partial_jf \sim \mu^2X^2,\end{equation}
the dimension parametrized by $X$ is frozen to $X=0$, as explained above.
  
To introduce spacetime dependence on some other coordinate, we just need to let $\mu$ above to depend on it, taking $\mu(X_9)$. The resulting ODE for $\mu$ is 
\begin{equation}\partial^2_9 \mu=0,\end{equation}
so that having $\mu=\mu_0\, X_9$ leads still to a classically marginal tachyon perturbation. The corresponding tachyon, which we now write in full detail,
 \begin{equation} \mathcal{T}= X_9\,( X^2 + \frac{\sqrt{2}}{q\beta}  X^++c)\, \exp(\beta X^+) , \quad q\beta\equiv \frac{\sqrt{2}}{\alpha'},\label{ewe}\end{equation}
 describes the backreaction close to a zero of the tachyon field between two regions where the tachyon acquires a very large vev. In these regions, the tachyon condenses to 10d type 0A, 0B respectively.

Let us spell out the effects of the superpotential interaction implied by the tachyon profile (\ref{tachyon-kink-type0}). Using (\ref{wsupo}), the resulting 2d scalar potential and fermion Yukawa coupling is
\bea
\Delta{\cal L}\sim (2X_9X)^2+(X^2)^2+2X_9 {\tilde \psi}\psi + 2X ({\tilde\psi}\psi_9+\psi{\tilde \psi}_9)\, .
\label{type0-couplings}
\eea
The remaining spacetime after tachyon condensation is given by the locus of vanishing scalar potential. Note that the second term forces $X=0$, so the first vanishes even for arbitrary $X_9$. Hence the resulting string theory propagates in 10d, with the field $X$ being gapped for generic $X_9$, but becoming gapless at $X_9=0$ (as explained in  \ref{sec:quantum}, this theory is non-compact in the technical sense of \cite{Tachikawa:2025awi}). 
At this $X=0$ slice, the fermion couplings reduce to a mass term for the 11th fermion, with a sign that flips as we move from negative to positive $X_9$. As already explained, this leads to a flip in the relative worldsheet fermion numbers of left- vs. right-moving Ramond groundstates, and hence of the GSO projection (formally, the worldsheets on the two sides of $X_9=0$ differ in the stacking of the Arf TQFT). The $X_9=0$ locus therefore defines a cobordism domain wall separating 10d type 0A and 0B theories. 

Let us emphasize again that, although our description above is in terms of the supercritical interpretation of the extra multiplet $X$, it is possible to consider the above construction purely from the viewpoint of considering the worldsheet theory as describing configurations of critical 10d strings, with extra gapped degrees of freedom that we integrate in or out to provide a non-trivial interpolation between the type 0A and 0B theories as we cross the locus where the gap closes. This viewpoint aligns with the discussion in section \ref{sec:obstructions} and is useful for examples where there is no supercritical string interpretation, or if we want to prioritize the critical string interpretation.

\subsection{More worldsheet physics: Majorana zero mode and the fate of D-branes}
\label{sec:type0-more}

The worldsheet theory we have constructed thus describes the dynamics of a fundamental string stretching across the 0A/0B domain wall, along which the GSO projection changes. Therefore, it is a type 0 cousin of the fundamental string stretching across the type IIA/IIB domain wall introduced in \cite{Heckman:2025wqd}. Indeed, in the following, we recover (the type 0 avatar of) several worldsheet properties considered there.

\subsubsection{The 1d Majorana fermion}
\label{1dfermion}

A well-established fact about the Arf TQFT is that, when defined on a 2d manifold with boundary, the boundary supports an edge mode given by a 1d Majorana fermion. This is required because the Arf invariant has a $\Z_2$ anomaly localized on the boundary \cite{Kitaev:2000nmw,Karch:2019lnn,Kaidi:2019pzj,Kaidi:2019tyf,Witten:2023snr,Freed:2024apc}, which is precisely canceled by introducing a localized 1d Majorana fermion (more succinctly, the Arf TQFT is the anomaly theory of this 1d fermion). Hence, such 1d fermion should arise on the worldsheet of a string stretched across a GSO domain wall, at the 1d interface where the worldsheet meets the wall. This was already pointed out in \cite{Heckman:2025wqd} (for IIA/IIB domain walls) at the topological level, but the lack of a microscopic description prevented a first principle derivation of this fact.

In contrast, our worldsheet description of the type 0A/0B domain wall does provide a simple microscopic derivation of the appearance of this 1d Majorana fermion. If we consider the worldsheet of the string stretching across the domain wall, the motion in $X_9$ corresponds to motion along the worldsheet coordinate $\sigma$. The 2d fermions ${\tilde \psi}$, $\psi$ are then a 2d Majorana fermion with a $\sigma$-dependent mass term with a simple zero at the value corresponding to $X_9(\sigma)=0$. Solving the equation of motion for the 2d fermion shows that there is a normalizable fermion zero mode localized on the interface. At the topologial level, this is precisely the 1d Majorana fermion at the boundary of the domain of the Arf TQFT. Our construction hence not only reproduces the required topological ingredients, but also provides a microscopic description for them.

\subsubsection{D-branes across the domain wall}
\label{sec:d-branes}

We can now use the above results to derive the effects experienced by D-branes that stretch across the domain wall, recovering the type 0 avatar of the results in \cite{Heckman:2025wqd} for the type IIA/IIB domain wall.

The spectrum of D-branes of the 10d type 0 theories can be simply described as a doubled version of the corresponding 10d type II theories (see e.g. \cite{Klebanov:1998yya,Bergman:1997rf,Billo:1999nf}). 
There are two sets of D$p$-branes, with $p$ even for type 0A and odd for type 0B, charged under the two RR $(p+1)$-form potentials. In addition, there are two sets of uncharged D$p$-branes (which we denote by ${\widehat{{\rm D}p}}$-branes), with $p$ odd for type 0A and even for 0B. The latter can be characterized by the fact that an open string tachyon kink on them produces, via open tachyon condensation, the corresponding charged D$(p-1)$-brane. These are the type 0 avatars of the BPS and unstable non-BPS D-branes in type II theories, and we refer to them as charged and uncharged in the type 0 context.
As is familiar from the construction of unstable non-BPS D-branes in type II theories, \cite{Sen:1998tt,Sen:1998ki,Witten:1998cd,Sen:1999mg} (see also  \cite{Kaidi:2019pzj,Kaidi:2019tyf,Witten:2023snr} for recent developments) ${\widehat{{\rm D}p}}$-branes are described essentially by the same boundary conditions as D$p$-branes, with the addition of one extra fermion on the 1d worldsheet boundary.

\begin{figure}[htb]
\begin{center}
\includegraphics[scale=.4]{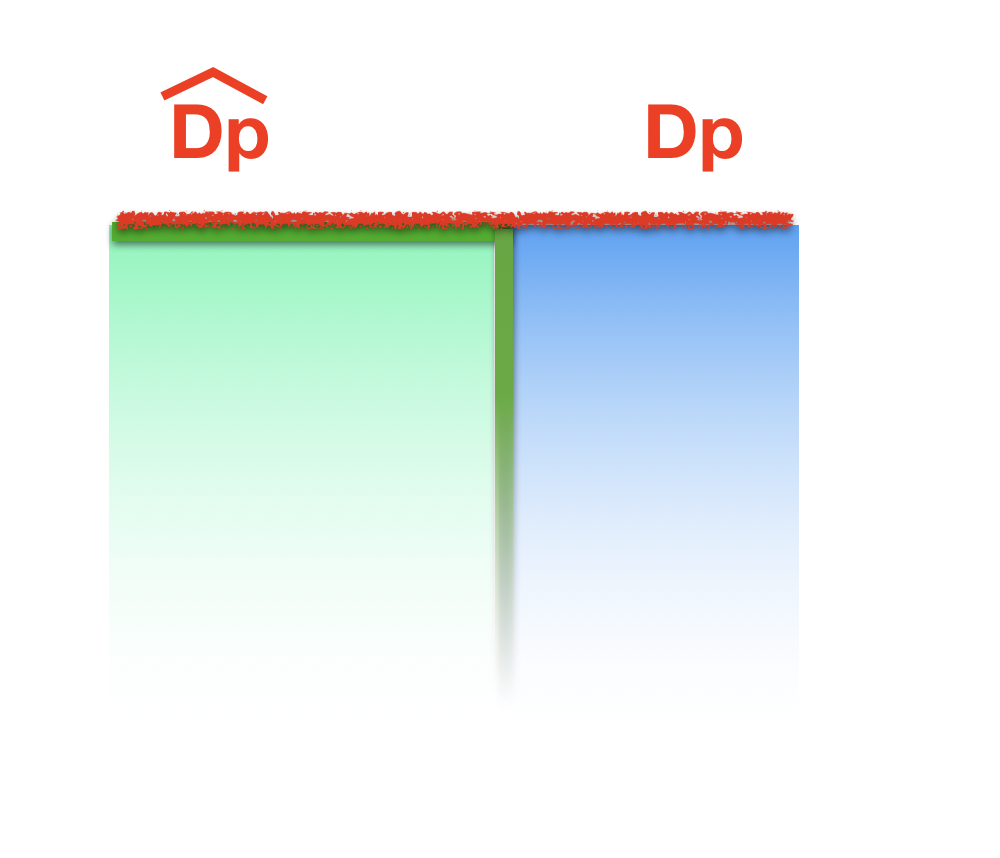}
\caption{\small Worldsheet for a string ending on a D-brane crossing the type 0A/0B domain wall. The worldsheet boundary is shown as the horizontal red line. The 2d green area corresponds to the theory with the stacked Arf TQFT, and has a 1d Majorana fermion localized on its boundary (green line), which includes the interface with the blue area as well as half of the worldsheet boundary. The presence of an extra fermion implies that a D$p$-brane on one side transforms into a ${\widehat{{\rm D}p}}$-brane on the other, and vice versa.}
\label{fig:worldsheet-bdry}
\end{center}
\end{figure}

Let us now exploit those ingredients to explain the behavior of D-branes upon crossing the type 0A/0B domain wall constructed in terms of our microscopic worldsheet description. Consider a 2d worldsheet with a boundary on a D-brane which crosses the domain wall, see Figure \ref{fig:worldsheet-bdry}. The blue and green shaded areas correspond to the two regions on the worldsheet, separated by 1d interface, shown as a green vertical line. The green area has the stacked Arf theory, so there is a 1d Majorana fermion on its boundary, which has two pieces: the first corresponds to the 1d interface between the two type 0 worldsheet regions, while the second corresponds to the boundary of the green area on the corresponding D-brane. The presence of this extra 1d fermion mode on the worldsheet boundary provides the key to understand the fate of D-branes upon crossing the domain wall. Indeed, the worldsheet boundary of a charged D$p$-brane of the type 0 theory on the right side acquires an additional fermion zero mode upon crossing the domain wall, turning it into the worldsheet boundary of an uncharged ${\widehat{{\rm D}p}}$-brane on the left side. Conversely, the worldsheet boundary of an uncharged ${\widehat{{\rm D}p}}$-brane on the right side already comes with one extra 1d fermion zero mode, so upon crossing the domain wall it acquires another one, and both may be gapped by a 1d mass term (corresponding to the fact that the relevant fermion anomalies are $\Z_2$-valued), leaving a worldsheet boundary with no extra fermions, namely we have a charged D$p$-brane of the left side.

%%%%%%%%%%%
%%%%%%%%%%%

The behavior of D-branes upon crossing the domain wall should match the behavior of RR $p$-forms upon crossing explained in \cite{Heckman:2025wqd}, namely the $p$-form potentials of the type 0 theory turn into the $p$-form field strengths of the other. This is directly discussed in appendix \ref{app:rrforms} via a topological spacetime action for the configuration.

\subsection{Quantum corrections and the strong coupling region}
\label{sec:quantum}

So far, it seems that we have succeded in attaining a worldsheet description of the 0A/0B domain wall. This may be regarded as an honest description of the domain wall in supercritical string theory, or merely as an off-shell  explicit description of a domain wall in the critical type 0 theory, which follows an up-and-down RG flow between the 0A/0B. The discussion in Section \ref{sec:obstructions} implies that the domain wall described above must involve a non-compact CFT at its core, in order to avoid the elliptic genus obstruction described there. We will now see that this is the case. The precise way to do this is explained in \cite{Tachikawa:2025awi}, and amounts to taking the superpotential from \eq{tachyon-kink-type0}, and adding a mass term and a vev $x_9$ for the coordinate $X_9$. Thus, we replace the superpotential from \eq{tachyon-kink-type0} by
\begin{equation}W=\frac12 X_9\, X^2 +\frac12m^2(X_9-x_9)^2.\label{pupu5}\end{equation}
The role of the mass term, which we ultimately remove by sending $m\rightarrow\infty$, is to set $\langle X_9\rangle=x_9$, and provides a precise meaning to the notion of looking at the up-and-down RG flow at a particular slice in the bordism. If the theory is compact for all $x_9$, then we say the theory with superpotential \eq{tachyon-kink-type0} is mildly non-compact; as shown in \cite{Tachikawa:2025awi}, this is the worldsheet analog of ``bordism with compact manifolds only'', and it means that the elliptic genus and \emph{tmf} classes are invariant under bordisms. 

The scalar potential coming from \eq{pupu5} is
\begin{equation} V= X_9^2X^2+ [\frac12X^2+m^2(X_9-x_9)]^2.\end{equation}
We wish to take the $m\rightarrow\infty$ limit. To do this, define a new scalar field
\begin{equation}\phi\equiv X_9-x_9+\frac12\frac{X^2}{m^2}.\end{equation}
In terms of these, the scalar potential takes the form
\begin{equation} V=\left(\phi+x_9-\frac12\frac{X^2}{m^2}\right)^2 X^2  + m^4 \phi^2, \end{equation}
and the kinetic term for $\phi$ is non-canonical,
\begin{equation} dX_9^2+ dX^2=\left(d\phi - \frac{X dX}{m^2}\right)^2+ dX^2. \end{equation}
As we send $m\rightarrow\infty$, all terms suppressed by powers of $m$ disappear, and only the  $m^4\phi^2$ term survives in the scalar potential, enforcing $\phi=0$. The resulting theory has a scalar potential
\begin{equation}V=x_9^2X^2\, ,\end{equation}
which is noncompact for $x_9=0$. As a result, the superpotential \eq{pupu5} is \emph{not} a mildly non-compact bordism, and we will be able to connect theories like 0A and 0B.
As explained in Section \ref{sec:obstructions}, it is natural to expect that a non-compact bordism like this will lead to some surprising properties for the domain wall. In our case, we can determine the IR limit of the RG flow triggered by the superpotential \eq{tachyon-kink-type0} (or more precisely, \eq{ewe}) explicitly. We will do this in two steps: first integrating out the supercritical coordinate $X$, leaving an effective field theory depending only on $X_9$ and $X^+$, and then studying the dynamics of the latter. Furthermore, since we are following an RG flow, we expect the UV and IR central charges to differ. To ensure that the IR central charge matches the value of a critical string theory, we will take the parameter $q$ in \eq{ewe} (which fixes the linear dilaton piece of the central charge) as free, and adjust its value at the end of the calculation.

Classically, the equation of motion for $X$ sets that coordinate to zero; so treating $X^+$ as a background field for the moment, we expect to have a flat direction parametrized by $X_9$. At non-zero $X_9$, $X$ is simply a massive field whose mass is controlled by $X_9$.  As explained in \cite{Witten:1995em,Hellerman:2006ff,Hamada:2021yxy}, integrating out $X$ leads to a one-loop renormalization of the kinetic term of $X_9$ and the dilaton profile, of the form \cite{Hellerman:2006ff}
\begin{equation}\delta G_{\mu\nu} =  \frac{\alpha'}{4} \frac{\partial_\mu M \partial_\nu M}{M^2},\quad \Delta\Phi = \frac14\ln\left(\frac{M}{\tilde{\mu}}\right)\, ,\end{equation}
where 
\begin{equation}M= 2X_9\, e^{\beta X^+}\, ,\end{equation}
is the mass of $X$, $\tilde{\mu}$ is a renormalization scale (that can be set by the value of the dilaton at any given point) and we have taken into account that a $(1,1)$ multiplet contains a Dirac fermion and a free scalar. The renormalization of the light-like components of the metric is exactly as in \cite{Hellerman:2006ff}. The only renormalized quantities are
\begin{equation}  \delta G_{++} =\frac{\alpha'}{4}\beta^2\quad ,\quad  \delta G_{99} =\frac{\alpha'}{4}\frac{1}{X_9^2},\quad \delta G_{9+} =\frac{\alpha'}{4}\frac{\beta}{X_9},\quad \Delta\Phi= \frac{1}{4}\left(\beta\, X^+ + \ln(2X_9)\right).\end{equation}
This is qualitatively similar to the results of \cite{Hellerman:2006ff}, but there is an additional dependence on the $X_9$ coordinate. The change of coordinates \begin{align}
U &= X^+, \\
V &= X^{-}- \frac{\alpha \beta}{4}\ln X_9 - \frac{\alpha \beta^{2}}{8}\,X^{+} \\
Y_9 &= \int^{X_9}_{X_9^0} \sqrt{1 + \frac{\alpha}{4 \xi^{2}}}\, d\xi \, ,
\label{metricdef}\end{align}
brings the metric to the flat Minkowski form
\begin{equation}
ds^{2} = -2\, dU\, dV + dY_9^{\,2},
\end{equation}
Notice that the naive junction point $X_9=0$ has been sent to infinite distance. Since the $X_9>0$ and $X_9<0$ regions correspond to 0A/0B respectively, this means that what we have constructed is not a worldsheet domain wall between these theories, since both sides have been pushed to infinite distance from one another in string units. In these coordinates, the full linear dilaton profile is
\begin{equation}\Phi+\Delta\Phi= a\, U + b\, V + c\, \ln X_9,\end{equation}
where $a,b,c$ are polynomial functions of $\beta$, $q$, and $\alpha'$ whose exact expression can be determined but is not very illuminating. $X_9$ is a non-linear function of  the canonically normalized coordinate $Y_9$, which is obtained by inverting the third equation of \eq{metricdef}. 

At this level of description, therefore, the IR limit of the worldsheet domain wall we constructed is flat space supplemented by a non-linear dilaton background. Close to $X_9=0$, that is, for large $Y_9$, the relationship is approximately linear, since $Y_9\sim -\ln X_9$. We conclude that the region close to the domain wall is represented by a flat metric in string units and a linear dilaton background! This is a valid worldsheet CFT, provided that the constants $a,b,c$ are tuned to satisfy the spacetime equations of motion. In ten-dimensional string theory, these imply that the dilaton is \emph{lightlike}, i.e. its gradient is a null vector. 

\begin{figure}[!htb]
\begin{center}
\includegraphics[width=0.95\textwidth]{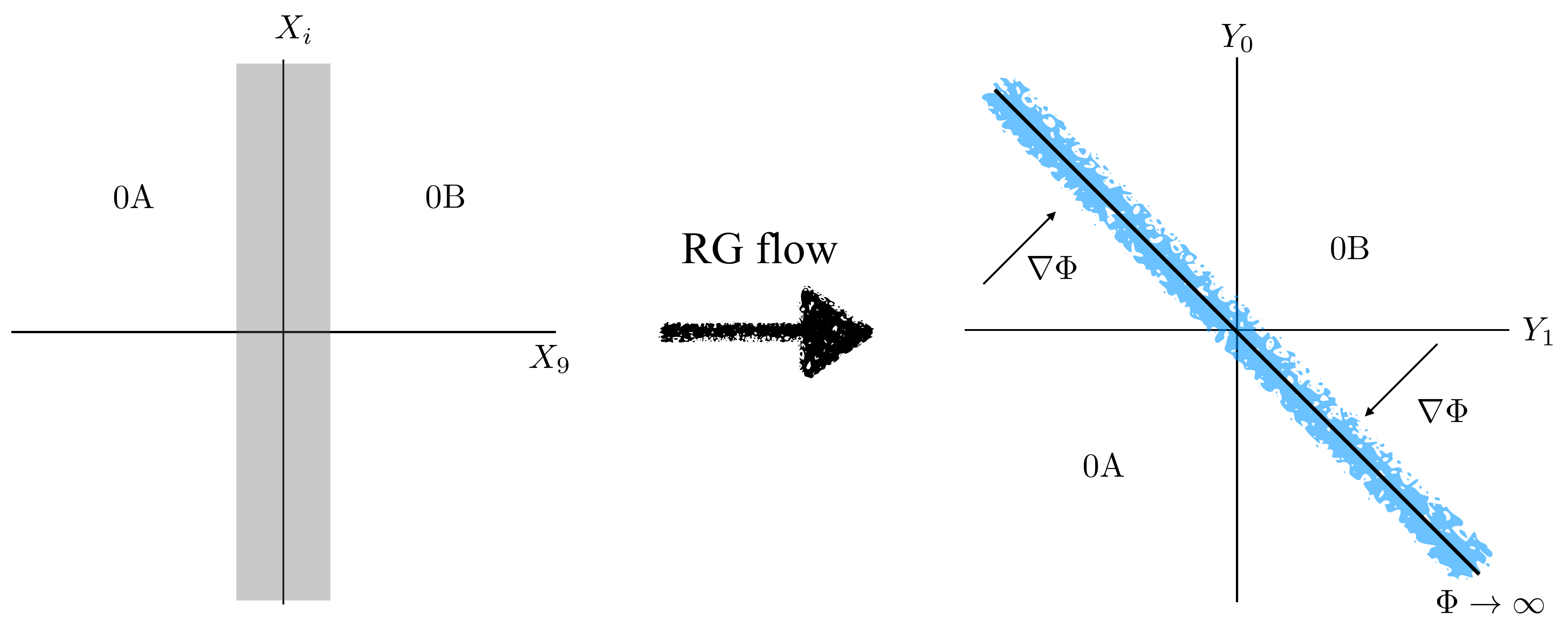}
\caption{\small On the left hand panel, we have a schematic depiction of the initial worldsheet QFT we start with, with tachyon profile as in \eq{ewe}. Away from the greyed-out $X_9=0$ region, the system flows to the 0A/0B vacuum. Close to $X_9=0$ we have a mildly non-compact CFT interpolating between the two solutions, but the string dilaton is small throughout the whole picture. On the right-hand panel, we have the actual IR CFT, which describes the actual spacetime profile corresponding to this configuration. The non-compact CFT has been replaced by two light-like linear dilaton backgrounds (the coordinates $Y_0,Y_1$ have a flat Minkowski metric and have been introduced for illustration) which are back-to-back, with the strongly coupled region of the 0A in the future next to the past strongly coupled region of 0B. We also depict the linear dilaton gradient on each side of the domain wall. The strongly coupled region in between the two means that the 0A/0B domain wall configuration goes beyond a perturbative worldsheet description.}
\label{fig:lindi}
\end{center}
\end{figure}

The light-like linear dilaton solutions of string theory are well understood \cite{Craps:2005wd}. Writing the dilaton profile as $\Phi\sim -X^+$ for some null coordinate $X^+$, they describe a weakly coupled region in the asymptotic $X^+\rightarrow\infty$ region, where the coupling is weak and strings propagate in a weakly coupled ten-dimensional spacetime (albeit with a dilaton gradient of stringy size, so the solutions are not described in supergravity), together with a strongly coupled region at $X^+\rightarrow-\infty$, where the dilaton diverges. The region of strong coupling $\Phi\sim 1$ is at finite distance, and corresponds to the location where the 0A/0B domain wall would be.  As a result, our attempt in producing a worldsheet description of the 0A/0B domain wall via a non-compact bordism has failed, in agreement with the general considerations of Section \ref{sec:obstructions} above; instead, we seem to have produced an object with a strongly coupled core, similar to the non-supersymmetric heterotic branes of \cite{Kaidi:2023tqo,Kaidi:2024cbx}. In contrast to these, our solution is intrinsically time-dependent, and describes at best a domain wall propagating at the speed of light, as illustrated in Figure \ref{fig:lindi}. Possibly, this feature arises due to the usual difficulties with constructing non-BPS brane solutions.

We also note in passing that the linear dependence of $X_9$ in \eq{ewe} is immaterial. The calculations work the same for a general function $f(X_9)$, and the conclusions are identical provided that $f(X_9)$ is linear for small $X_9$.

Let us note that we could have guessed this linear dilaton result from general considerations. In ten dimensions, the domain wall is an object of codimension 1, and should therefore be described by a worldsheet CFT involving 8 scalars $X_1,\ldots X_8$ parameterizing the worldvolume directions, times a possibly non-trivial worldsheet CFT describing the time and transverse directions. Cancellation of the conformal anomaly against the ghosts implies that this extra factor has central charge $c=2$, and it involves the time-like direction. It should also describe some region, far away from the domain wall, where the configuration is somewhat close to flat, ten-dimensional space. In ordinary perturbative string theory, the only such background with a simple spacetime interpretation is precisely the light-like linear dilaton that we found.

We have therefore provided a description of the 0A/0B domain wall as complete as possible within perturbative string theory, and our construction moreover predicts the end of its regime of validity by producing a strongly coupled region. The key question that remains unaddressed is whether we have really succeeded in constructing a transmissible 0A/0B domain wall. This may be supported by how our construction reproduces (type 0 avatar of) the behavior in \cite{Heckman:2025wqd} of fields and objects (e.g. D-branes) across the domain wall, and that the core of the defect is (extending naively the solution admittedly beyond its regime of validity) at finite distance in the Einstein frame \cite{Angius:2022mgh}. On the other hand, it could be that the strongly coupled region does not allow for transmission of information and energy between the 0A/0B sides. It could also happen that the strongly coupled region hides an infinite region of space, so that the object that we have constructed is not a domain wall, but rather, two completely disconnected  non-compact 0A and 0B backgrounds. Answering either of these questions would require an understanding of the defect beyond perturbation theory, which we currently do not possess; we comment on some speculations in the conclusions.

\subsection{An aside: The heterotic GSO domain wall}
\label{sec:heterotic}

In this section we make a short aside to provide a worldsheet description for the cobordism domain wall between two 10d $Spin(32)/\Z_2$ theories which differ by the chirality of their massive spinor state, introduced in \cite{Heckman:2025wqd}.

We describe the 10d heterotic theory by a $(0,1)$ worldsheet with (ignoring ghosts) 10 spacetime bosons $X^\mu$, their 10 right-moving Majorana-Weyl fermion partners $\psi^\mu$, and 32 left-moving Majorana-Weyl spinors $\lambda^a$ in Fermi supermultiplets. There is a $\Z_2$ gauge symmetry given by overall worldsheet fermion number, and we further gauge a $\Z_2$ corresponding to e.g. left-moving worldsheet fermion number, leading to the $SO(32)$ spacetime gauge symmetry (the gauging of right-moving worldsheet fermion number, producing the supersymmetric right-moving GSO projection, is simply a combination of the two previous $\Z_2$'s). Equivalently, we sum over independent spin structures for left- and right-movers. Hence, one can build two theories that  differ
by the stacking of an Arf TQFT on e.g. the left-moving sector, so that the GSO projection is flipped on its Ramond groundstates. The two resulting theories are almost identical and differ only in the chirality of massive spinor states.

It is straightforward to propose a one-parameter family of CFT's interpolating between the two theories. We simply realize  that again the stacking (or not) of the Arf TQFT can be reproduced by the introduction of an additional Majorana fermion (whose left- and right-moving parts we denote by ${\tilde \lambda}$, $\psi$, respectively) with a mass term whose sign flips as we change with the interpolating parameter. The extra degrees of freedom and interactions can be made manifestly $(0,1)$ by including a bosonic partner $X$, so that the mass terms arise from the $(0,1)$ superpotential
\bea
{\Delta {\cal L}}\sim \mu\int d\theta_+  {\tilde \lambda} X\, ,
\label{supo-het}
\eea 
which makes the boson acquire a mass squared $\mu^2$ and the fermion a mass $\mu$, which flips sign at $\mu=0$. As in the type 0 case, the theory is gapped for generic $\mu$, but the gap closes at $\mu=0$, where it develops a non-compact direction which allows for the interpolation between two topological different theories.

The above family can be given an explicit physical realization by using a supercritical heterotic string construction, which is reviewed in appendix \ref{app:heterotic}. The worldsheet theory of the 11d supercritical heterotic string has (ignoring ghosts) 11 bosonic coordinates $X^\mu$, $X$ (with $\mu=0,\ldots, 9$), 11 right-moving fermion partners $\psi^\mu$, $\psi$, and 33 left-moving fermions ${\tilde \lambda}^a$, ${\tilde\lambda}$ (with $a=1,\ldots, 32$). There is a gauged $\Z_2$ corresponding to overall worldsheet fermion number. We focus on a particular heterotic theory, studied in \cite{Hellerman:2006ff}, with a further gauging of a $\Z_2$ generated by $(-1)^{f}R$, where $R$ flips the extra coordinate $X$ and $(-1)^{f}$ flips the extra right-moving fermions $\psi$,  and the 33 left-moving ones. The resulting theory has an 11d untwisted sector containing a closed tachyon, the graviton, 2-form, dilaton, and $SO(33)$ gauge bosons, and a twisted sector localized on the 10d slice $X=0$ containing fields ultimately corresponding to the gravitino, dilatino and gauginos. Tachyon condensation is described by the introduction of a $(0,1)$ superpotential, so that the expression (\ref{supo-het}) corresponds to (once dressed with a suitable lightlike exponential in $X^+$) a marginal deformation removing the extra dimension and bringing the theory down to 10d heterotic $Spin(32)/\Z_2$ theory. As in the type 0 case, the sign of $\mu$ determines which of the two 10d $Spin(32)/\Z_2$ heterotic theories is recovered, namely flips the GSO projection for the left-moving Ramond groundstate.

It is now straightforward to give a worldsheet description for a cobordism domain wall between these two 10d heterotic theories. We simply define a tachyon kink, by promoting the parameter $\mu$ in the previous discussion to an explicit dependence on a 10d spacetime coordinate, say $X_9$. At leading order, this gives the superpotential
\bea 
\Delta{\cal L}=\int d\theta_+ {\tilde \lambda}\,e^{\beta X^+} X_9 X\, ,
\eea 
where for completeness we have restored the lightlike exponential.
Using (\ref{heterotic-tachyon-condensation}), at late $X^+$ one obtains the scalar potential and fermion couplings
\bea 
\Delta{\cal L}\sim X_9^2 X^2 + X_9{\tilde \lambda}\psi\, .
\eea 
For generic $X_9$ the boson $X$ and the fermions ${\tilde \lambda}$, $\psi$ are massive, with the mass of the latter flipping sign across $X_9=0$, and reproducing the stacking of the Arf TQFT on one of the sides of the wall. At $X_9=0$, the extra dimension $X$ opens up as befits the non-compact theory required for the interpolation between two topologically different theories.  

A similar calculation to the one in the previous subsection shows that this worldsheet QFT is also mildly non-compact in the sense of \cite{Tachikawa:2025awi}. In any event, at the quantum level, we recover the same behavior, with the point $X_9=0$ being driven to infinite distance, corresponding to a strong coupling region at the core of the wall. In fact, the combination of the dilaton backgrounds results in the two sides of the domain wall developing a light-like linear dilaton. Assuming these are resolved as end of the world boundaries, the domain wall is ultimately described as two back-to-back big bang/big crunch bubbles of nothing. However, as in the type 0 case, we remain agnostic in what respects the non-perturbative resolution of the strong coupling singularity. We again emphasize that we have attained a description of the heterotic GSO domain wall which is as good as it can get within the realm of perturbative string theory. In particular, it predicts the breakdown of its own regime of validity, in agreement with the arguments in Section \ref{sec:obstructions}.

We refrain from further discussing the physics of this heterotic domain wall, and turn to the discussion of the more interesting case of the type IIA/IIB domain wall.

\section{Type IIA/IIB domain wall from worldsheet cobordisms}
\label{sec:typeII}

In this section we explore the extension of the constructions in the previous section to type II theories and discuss the description of the type IIA/IIB cobordism domain wall via worldsheet interpolations involving non-compact CFTs. The extension requires the introduction of several interesting new ingredients.

\subsection{The worldsheet interpolation and symmetric mass generation}
\label{sec:smg}

We start by considering the construction of an interpolation between the worldsheet theories of 10d type IIA and IIB theories. In analogy with the type 0 setup, both theories differ in the stacking of an Arf TQFT, which flips the GSO projection on the Ramond groundstate on e.g. the left-moving sector. This is equivalent to the statement that the type II theories are obtained from the corresponding type 0 theories by gauging the $\Z_2$ symmetry generated by $(-1)^{F_L^{ws}}$, with $F_L^{ws}$ being left-moving worldsheet fermion number.

This suggests that in order to obtain an interpolation between the worldsheet theories, we must naively include additional fermion degrees of freedom with a mass term whose sign changes in the interpolation, thus providing the required appearance of the Arf TQFT. However, the possibilities of adding extra fermions are constrained by the $\Z_8$ anomaly\footnote{Actually, the obstruction for the theory to admit a modular invariant partition function defines a $\Z_{16}$ anomaly \cite{BoyleSmith:2024qgx,Heckman:2025wqd}, so if we add just 8 fermion fields, we will change the worldsheet theory from fermionic to bosonic, when performing the GSO projection. We can go around issue by simply introducing two copies of these 8-plets of extra fields in the discussion below, but at any rate, this only affects the spins of the R groundstates, which may change from integer to half-integer worldsheet spin.\label{foot:backfire} } of 2d fermion theories invariant under the chiral $\Z_2$ generated by $(-1)^{F_L^{ws}}$. In order to cancel it, we must include a set of 8 Majorana fermions, whose left- and right-moving components we denote by ${\tilde \chi}^i$, $\chi^i$ respectively, with the former charged under $(-1)^{F_L^{ws}}$. To construct a valid interpolation, the extra anomaly-free degrees of freedom should be gapped at generic values of the interpolating parameter, but this cannot be done by a naive mass term, because it is not invariant under $(-1)^{F_L^{ws}}$. Hence, we must consider mechanisms to gap the anomaly-free set of 8 fermions in a way compatible with the $\Z_2$ symmetry, which forbids a naive mass term. These mechanisms are known as symmetric mass generation (see e.g. \cite{Wang:2022ucy} for a review), and have been actively studied in particular in 2d. In fact, the system of 8 Majorana fermions is a paradigmatic and well studied example, from different viewpoints \cite{Fidkowski:2009dba,Ryu:2012he,Qi:2012gjs,Tong:2019bbk}, which we now review following \cite{Tong:2019bbk} (to which we refer the reader for further details).

Consider a 2d theory of 8 Majorana fermions, with ${\tilde\chi}_i$, $\chi_i$ as left- and right-moving components. The system has an anomaly-free $\Z_2^V\times\Z_2^A$ symmetry, associated to the overall fermion number, and left-moving fermion number, respectively. Although the naive mass term is forbidden by the symmetry, it is possible to gap the set of 8 fermions while preserving the $\Z_2^V\times\Z_2^A$ symmetry via symmetric mass generation, as follows. The crucial ingredient is that the set of fermions transforms in the ${\bf 8_v}$ representations of the $Spin(8)$ symmetry, and this is related to the chiral spinor representations via triality ${\bf 8_v}\to{\bf 8_s}\to{\bf 8_c}\to {\bf 8_v}$. We refer the reader to \cite{Tong:2019bbk} for details, but the gist of the idea is that a set of 8 fermions in any of these representations, once suitably coupled to $\Z_2$ gauge fields, may be recast as a set of fermions in any other via a change of variables, which is most simply described as a change of basis in the bosonized version of the fermions. 

The symmetric mass generation for 8 Majorana fermions is described by representing them as fermions ${\tilde\chi}_i'$, ${\chi}_i'$ in the ${\bf 8_s}$, and considering an $Spin(7)$ invariant (rather than a full $Spin(8)$ invariant) 4-fermion interaction with the structure
\bea 
{\cal L}_{Spin(7)}=-A\left(\sum_{i=1}^7{\tilde \chi}_i'\chi_i'\right)^2-B \left(\sum_{i=1}^7{\tilde \chi}_i'\chi_i'\right){\tilde \chi}_8'\chi_8'\, .
\label{spin7-int}
\eea 
In the regime $|A|\gg |B|$, the dynamics is dominated by the first term, which corresponds to an $Spin(7)$ Gross-Neveu model \cite{Gross:1974jv}, which develops a bilinear condensate
\bea
\langle \;\sum_{i=1}^7{\tilde \chi}_i'\chi_i'\;\rangle=\Lambda\, .
\eea 
The 7 Majorana fermions get a mass of order $\sim |A|\Lambda$ and can be integrated out. We are left with an effective theory with only one lighter Majorana fermion ${\tilde \chi}_8'$, $\chi_8'$ with an effective mass term $m\sim -B\Lambda$ (plus topological couplings). The 8 Majorana fermions are therefore ultimately gapped, completing the symmetric mass generation for this system. For later discussion, we emphasize this interesting intermediate effective description with a single Majorana fermion with a mass term (in a way compatible with the $\Z_2^V\times \Z_2^A$ symmetry).

This allows the following interpolation between the 10d type IIA and type IIB worldsheet theories. Starting from one 10d type II theory, we focus on its fermion sector, which has a $\Z_2^V\times\Z_2^A$ gauge symmetry. We then add 8 extra Majorana fermions in the ${\bf 8_s}$ and turn on the $Spin(7)$ interactions (\ref{spin7-int}), to gap them in a way compatible with the $\Z_2^V\times\Z_2^A$ symmetry. We let the parameters, in particular the coefficient $B$, of the interaction change, so that the mass of the single Majorana fermion in the effective description flips sign, allowing for the appearance of a stacked Arf TQFT once all of the 8 fermions are integrated out.  The result is therefore a 10d theory with the opposite GSO projection for its left-moving Ramond groundstate. Since the whole process is compatible with the $\Z_2^V\times\Z_2^A$ symmetry, it is inherited in the quotient that turns the theory into a type II setup. The whole process corresponds to starting in a 10d type II theory, integrating in a set of fields, moving in parameter space, and integrating them out to land in the other type II theory. This is in the spirit of moving up-and-down the RG flow and describes a worldsheet cobordism. The key to the non-trivial interpolation between topological sectors of the 2d theory is that the interpolation crosses a point where the gap closes, so a non-compact CFT is involved, as explained in section \ref{sec:cheat}.

It is interesting to emphasize that the intermediate effective description in terms of just one Majorana fermion is essentially identical to that used in the construction of the cobordism domain wall in type 0 theories in section \ref{sec:type0}. This basically implies that many ideas realized there will work similarly in our present context, such as the appearance of the 1d Majorana fermion in the worldsheet interface, and the behavior of D-branes upon crossing the wall, hence recovering several key results of \cite{Heckman:2025wqd}. 

The above discussion of fermions certainly includes the right topological ingredients we need, but is not a complete worldsheet description yet. Actually, to have a real interpolation between type IIA and IIB theories, we need a $(1,1)$ supersymmetric version of the above mechanism, which we study in the next section.

\subsection{Supersymmetric completion of the worldsheet interpolation}
\label{sec:susy-smg}

In the above discussion we have focused on the physics of the worldsheet fermion sector. On the other hand, the actual realization of the interpolation for type II theories requires a $(1,1)$ version of the above mechanism. In the following we provide the basic ingredients for this description, with extra details provided in appendix \ref{app:one-comma-one}.

We would like to obtain a $(1,1)$ supersymmetric theory whose infrared dynamics is that of the purely fermionic theory in the previous section. As discussed in appendix \ref{app:one-comma-one}, a standard trick is to first linearize the quartic fermion interactions as Yukawa couplings with additional bosonic fields, and then promote the original fermions and the extra bosons to full supermultiplets. In order to apply this to the $Spin(7)$ interactions (\ref{spin7-int}), we write them as a complete square
\bea
{\cal L}_{Spin(7)}=-A\left(\sum_{i=1}^7{\tilde \chi}_i'\chi_i'+\frac B{2A} {\tilde \chi}_8'\chi_8'\right)^2\, .
\label{spin7-bis}
\eea 
where the quartic term in the Grassman variables ${\tilde \chi}_8',\chi_8'$ is identically zero. This works like a deformed version of the Gross-Neveu model in Appendix \ref{app:one-comma-one}.
We now introduce a (known as Hubbard-Stratonovich) scalar field $\sigma$ to linearize the fermionic interactions. 
\bea 
\mathcal{L}_{\rm lin} =  \frac{\sigma^2}{4A}  + \sigma\left( \sum_{i=1}^{7} {\tilde \chi}_i' \chi_i'
+  \frac{B}{2A}  {\tilde \chi}_8' \chi_8' \right)\, .
\eea 
The theory can be made $(1,1)$ supersymmetric by promoting the fermions $\chi_i',\chi_8'$ to supermultiplets $\Phi_i, \Phi_8$ and the boson $\sigma$ to a supermultiplet $\Sigma$ with superpotential
\bea 
W=\frac{1}{4A} \Sigma^2+ \left( \sum_{i=1}^{7} \Phi_i\Phi_i + \frac{B}{2A}\Phi_8\Phi_8\right) \Sigma \, .
\label{susy-gnh}
\eea
It is possible to show, by integrating out all fields except for the original fermions as in appendix \ref{app:one-comma-one}, that this $(1,1)$ theory leads to the same infrared dynamics as the purely fermionic model (\ref{spin7-int}). Hence this provides a $(1,1)$ version of the mechanism used in the previous section for the interpolation.

An amusing alternative description is obtained if we integrate out the multiplet $\Sigma$ in (\ref{susy-gnh}) and obtain
\bea
W=-A\left(\sum_{i=1}^7 \Phi_i^2+\frac B{2A} \Phi_8^2 \right)^2\, ,
\label{sgn}
\eea 
which is just the naive promotion of the fermionic lagrangian (\ref{spin7-bis}) to supermultiplets.

It is also possible to make more manifest the intermediate step with only one extra multiplet. For this, we consider the hierarchical regime of couplings $|B|\ll |A|$, in which the dynamics of the supersymmetric $Spin(7)$ Gross-Neveu-Higgs dominates, so that we have the strong dynamics condensate $\Lambda$. This can be done in either of the two previous formulations, for instance in (\ref{susy-gnh}). One can integrate out the supermultiplets $\Phi_i$ and $\Sigma$ to get an effective superpotential
\bea
W_{eff}=- B\Lambda \Phi_8^2+\ldots\, .
\label{effective-typeii}
\eea 
where the dots indicate extra terms, which can be ignored.
This is the supersymmetric completion of the effective description with a single extra fermion described in section \ref{sec:smg}, so the sign of the effective mass term for the supermultiplet (i.e. the sign of $B$) determines the appearance or not of the Arf TQFT in the resulting 10d type II theory (once the quotient by the chiral $\mathbb{Z}_2^A$ is taken).

Hence we have found a family of 2d $(1,1)$ worldsheet theories, depending on a parameter $B$, which interpolate between the 10d type IIA and type IIB worldsheet theories, at the price of including a non-compact theory for some value of the parameter in the interpolation. We now turn to the discussion of the dynamical implementation of this interpolation in spacetime, in order to describe the type IIA/IIB domain wall.

\subsection{Worldsheet description of the 10d type IIA/IIB domain wall}
\label{sec:typeii-wall}

In order to describe the physical realization of the above worldsheet cobordism as a spacetime type IIA/IIB domain wall, in analogy with the type 0 case, we should promote the interpolating parameter (in this case $B$) to a dynamical coordinate $X_9$. This is most easily done at the level of the theory (\ref{sgn}), leading to
\bea
W= -A\left(\sum_{i=1}^7 \Phi_i^2\right)^2-X_9 \left(\sum_{i=1}^7\Phi_i^2\right)\Phi_8^2\, .
\label{almost-supercritical-typeii}
\eea 
Here, as in the type 0 case, the linear dependence in $X_9$ should be regarded as just the leading approximation to an in principle more general function with a simple zero at $X_9=0$. A global choice leading to effective constant coefficient $B_{eff}$ with opposite signs as $X_9\to\pm \infty$ would be obtained by the replacement $X_9\sim B_{eff}\arctan X_9$.

At the level of the effective description (\ref{effective-typeii}), the domain wall would be given by 
\bea
W_{eff}=-\Lambda X_9 \Phi_8^2\, ,
\label{almost-supercritical-typeii-eff}
\eea 
where we again work in the spirit that the linear dependence in $X_9$ should be regarded as an approximation near $X_9=0$. 

This expression is very similar to the worldsheet interpolation description of the type 0A/0B domain wall in section \ref{sec:type0}, in particular the tachyon background (\ref{tachyon-kink-type0}) in the supercritical description of the type 0A/0B domain wall. Hence, the spacetime dependent worldsheet interpolation we have provided for the 10d type IIA/IIB domain wall is completely analogous to that of the 10d type 0A/0B domain wall. 

This ensures that we can use analogous arguments to the type 0 domain wall, and reproduce the properties of the type IIA/IIB domain wall in \cite{Heckman:2025wqd} essentially as discussed in section \ref{sec:type0} in the type 0 context. The (dis)appearance of the Arf TQFT upon integrating out heavy worldsheet modes follows just from the flip of the sign of the parameter controlling the (effective) fermion mass as we move across $X_9=0$. The 1d Majorana fermion on the 1d interface on a worldsheet stretching across the domain wall arises, as in section \ref{1dfermion}, as a fermion zero mode of the effective 2d Majorana fermion coupled to the boson $X_9$ as it crosses a simple zero. This again implies that, in the presence of worldsheet boundaries crossing the line defect (i.e. D-branes crossing the domain wall), boundary states pick up (or lose) an extra 1d fermion zero mode, as in section \ref{sec:d-branes}. Hence BPS D-branes in one 10d type II theory turn into non-BPS D-branes in the other theory, and viceversa. This effect dovetails the transformation of RR fields across the wall, given by restricting the description in Appendix \ref{app:rrforms} to the type II RR fields. 

Finally, once we dress the above interactions with an exponential of $X^+$, the discussion in section \ref{sec:quantum} about the quantum corrections carries over to the present type II context, with similar conclusions. The corrections to the spacetime metric and dilaton turn the location of the domain wall into an infinite strongly coupled throat. The domain wall degenerates into two back-to-back lightlike linear dilaton singularities. Again, a natural interpretation is that the cobordism between the 10d type IIA and IIB theories corresponds to the stacking of two (coincident and lightlike) bubbles of nothing for the two theories, but the actual non-perturbative resolution of the singularity remains beyond the reach of worldsheet physics. We nevertheless emphasize that our description is as good as perturbative string theory can get, with an explicit worldsheet description which predicts even the limitations of its regime of validity, in agreement with the arguments in section \ref{sec:obstructions}.

\subsection{Towards a supercritical description?}
\label{sec:typeii-supercritical}

As we have discussed, several ingredients in the worldsheet construction of the type IIA/IIB domain wall are reminiscent of the supercritical interpretation of the type 0A/0B domain wall. However, there are some important observations regarding the possible interpretation of the type IIA/IIB worldsheet interpolation in terms of a physical supercritical type II.

One key observation is that in the type 0 context, the extra $(1,1)$ supermultiplet describes an additional spacetime coordinate, and there is a regime (when the tachyon background is small, which is physically realized at $X^+\to -\infty$) in which the additional boson is related to the 10d ones by 11d Poincar\'e invariance (only broken by the linear dilaton background). On the other hand, in the type II context, the extra bosons are partners of the extra fermions, which, as discussed in section \ref{sec:smg}, are in the spinor representation. Hence, they do not have a direct spacetime interpretation, as they do not mix with the 10d coordinates in a Poincar\'e symmetry. This prevents a direct interpretation of the $(1,1)$ superpotentials for type II theories, such as (\ref{almost-supercritical-typeii}) or (\ref{almost-supercritical-typeii-eff}) as tachyon backgrounds for a supercritical theory. 

There is actually a deeper reason for this fact, related to a notable absence of tachyons in certain type II supercritical strings. As described in appendix \ref{app:typeII} (see \cite{Hellerman:2007fc} for details), there are two main supercritical extensions of type II theories, which we now compare with our construction in turns. 

\medskip

{\bf Supercritical type II theories in $D=2+8k$ dimensions}

The first class of supercritical extensions of type II theories involves the addition of extra 8-plets of bosons and fermions in the vector representation of the Poincar\'e symmetry, so that they are anomaly-free under the $(-1)^{F_L^{ws}}$ chiral symmetry flipping the sign of left-moving fermions. These theories are hence defined in $D=2+8k$ spacetime dimensions, and were discussed in \cite{Hellerman:2007fc} in detail. More specifically there are two series of such theories, the {\em standard series}, defined in $D=10+16k$ dimensions, and the {\em Seiberg series} in $D=2+16k$ dimensions. The two series differ slightly in the form of their partition function, due to the $\Z_{16}$ obstruction to modular invariance mentioned in footnote \ref{foot:backfire}. The fermionic sector used in our type II worlsheet constructions (c.f. section \ref{sec:smg}) is related to the fermionic sector of these supercritical type II theories by triality on the 8-plets. On the other hand, the bosons in our construction belong to different representations compared with those of these supercritical theories (morally, sets of ${\bf 8}_s$ in our construction, vs sets of ${\bf 8}_v$ in the supercritical type II). 

The fact that our construction does not correspond exactly to these supercritical type II theories is actually a fortunate one. Indeed, as reviewed in appendix \ref{app:typeII} (see \cite{Hellerman:2007fc} for details), the supercritical type II theories in $D=2+8k$ dimensions are remarkably non-tachyonic. This means that they do not admit $(1,1)$ superpotential deformations turning them into the 10d type II theories. Equivalently, the tachyon present in the supercritical type 0 theory in $D=2+8k$ dimensions is projected out by the action of $(-1)^{F_L^{ws}}$. In our construction, thanks to the fact that our fields transform in the ${\bf 8}_s$ instead of the ${\bf 8}_v$, the $(1,1)$ superpotential couplings are allowed and can be used to connect them to the 10d type II theories, via symmetric mass generation. The price to pay is precisely that there is no direct supercritical spacetime interpretation of the theory with the extra fields.

\medskip

{\bf Orbifolds of type 0 supercritical theories in $D=10+2k$}

There is a second version of supercritical theories (see appendix \ref{app:typeII}, and \cite{Hellerman:2006ff,Hellerman:2007fc} for details) which do contain tachyons and admit superpotential deformations to the 10d type II theory. They are described as supercritical type 0 theories in $D=10+2k$ dimensions quotiented by a $\Z_2$ symmetry acting as a sign flip of $k$ of the extra dimensions times an action anticommuting with the left-moving supercharge. This construction can be related to the effective worldsheet theory of the type II theory with $(1,1)$ symmetric mass generation, c.f. (\ref{effective-typeii}), as follows. Consider starting with a supercritical type II theory with two additional 8-plets of fields $\Phi_i$, $\Phi_i'$ (in fact, as required by the above mentioned $\Z_{16}$ obstruction in building type II modular invariant partition functions), and reduce the dynamics of each set to one effective additional multiplet $\Phi_8$, $\Phi_8'$. The resulting theory is two copies of (\ref{effective-typeii}), which may be recast as the quadratic superpotential (\ref{supo-supercritical-typeii}) for two extra multiplets. The modular invariant partition function for the resulting theory can be regarded as that of the supercritical extension of type II theory constructed as an orbifold of 12d type 0. Again, a subtle point lies in the fact that the Lorentz representation of the additional fields is different in the supercritical theory in appendix \ref{app:typeII} and in our construction. 

\medskip

We conclude by pointing out a further possible avenue towards a fully supercritical physical realization of the worldsheet interpolation with additional 8-plets that we have proposed. One may consider the supercritical type II theories in $D=2+8k$, i.e. with extra 8-plets of fields in the ${\bf 8}_v$ of the corresponding $Spin(8)$ symmetries, and use triality to transform the fermions into 8-plets in the ${\bf 8}_s$, so that they are gappable via the mechanism in section \ref{sec:smg} and we can implement the IIA/IIB worldsheet interpolation we have described. This formulation would amount to a supercritical analogue of the Green-Schwarz formulation of 10d type II theories, with worldsheet bosons transforming as spacetime vectors, and worldsheet fermions transforming as spacetime spinors. 

Actually, there is an interesting fully $(1,1)$ worldsheet supersymmetric way to implement the above procedure in a more direct way, by writing the $Spin(7)$ preserving interaction (\ref{spin7-int}) directly in terms of the original Majorana fermions $\psi^a$ in the ${\bf 8}_v$. This was already done in the original paper \cite{Fidkowski:2009dba} as
\bea
{\cal L}_{Spin(7)}=\Phi_{abcd}\psi^a\psi^b\psi^c\psi^d
\eea
with $\Phi_{abcd}$ being the $Spin(7)$-preserving Cayley 4-form of $Spin(8)$. By triality with the description with fermions in the ${\bf 8}_s$, this interaction leads to the desired symmetric mass generation. It is possible to rewrite this interaction using a set of 7 Hubbard-Stratonovich scalars (in the vector representation of $Spin(7)$), by exploiting the decomposition of the Cayley 4-form as a quadratic contraction of 7 gamma matrices acting on the 8-dimensional spinor of $Spin(7)$. One may subsequently promote to a fully $(1,1)$ supersymmetric version of the mechanism, in analogy with our discussion in the previous sections. This description, in which the extra bosons and fermions are in the vector representation of the corresponding part of the Lorentz group, would provide a full-fledged worldsheet description of the type IIA/IIB domain wall in a physical type II supercritical theory. However, the spacetime interpretation of the background corresponding to turning on/off the 4-fermion interaction remains challenging, and we leave it as an interesting open question for the future.

\subsection{Quantum corrections and the strong coupling region}
\label{sec:strong-typeii}

In analogy with section \ref{sec:quantum}, a more precise picture of the IIA/IIB domain wall configuration arises upon integrating out the massive degrees of freedom. The result of this computation is analogous to that in 0A/0B domain wall, as should be clear from using the intermediate effective picture with only one extra multiplet, essentially identical to the type 0 tachyon kink configuration. In addition, the general arguments in section \ref{sec:quantum} motivating that the only possible endpoint of the RG flow is two lightlike linear dilaton backgrounds applies to the type II case. Hence the actual configuration described by our worldsheet construction corresponds to a type II version of the left panel in Figure \ref{fig:lindi}, with a ligthlike strongly coupled core at which the worldsheet description fails, in agreement with the general arguments in section \ref{sec:obstructions}. 
We have therefore provided a description of the IIA/IIB domain wall as complete as possible within perturbative string theory, and our construction moreover predicts the end of its regime of validity by producing a strongly coupled region.

The key question is the nature of the resolution of this strongly coupled singularity, which we now discuss. As explained in \cite{Hellerman:2006ff} there is more than one UV completion to the linear dilaton background, since small perturbations in the small coupling region can have a large impact when coupling becomes large, so the physics is non-universal. Nevertheless, there are some well-known UV completions of the linear dilaton, such as the BPS one discussed in \cite{Craps:2005wd} in the case of IIA strings. The exact linear dilaton uplifts to a pure geometry BPS solution in M-theory which describes a ``matrix Big Bang'', or its CPT conjugate Big Crunch. If this behavior also occurs in the linear dilaton backgrounds we found, then the IIA/IIB domain wall we constructed is not transmissible, but corresponds to a IIA region that ends, followed by a separate IIB Universe.

\begin{figure}[!htb]
\begin{center}
\includegraphics[width=0.85\textwidth]{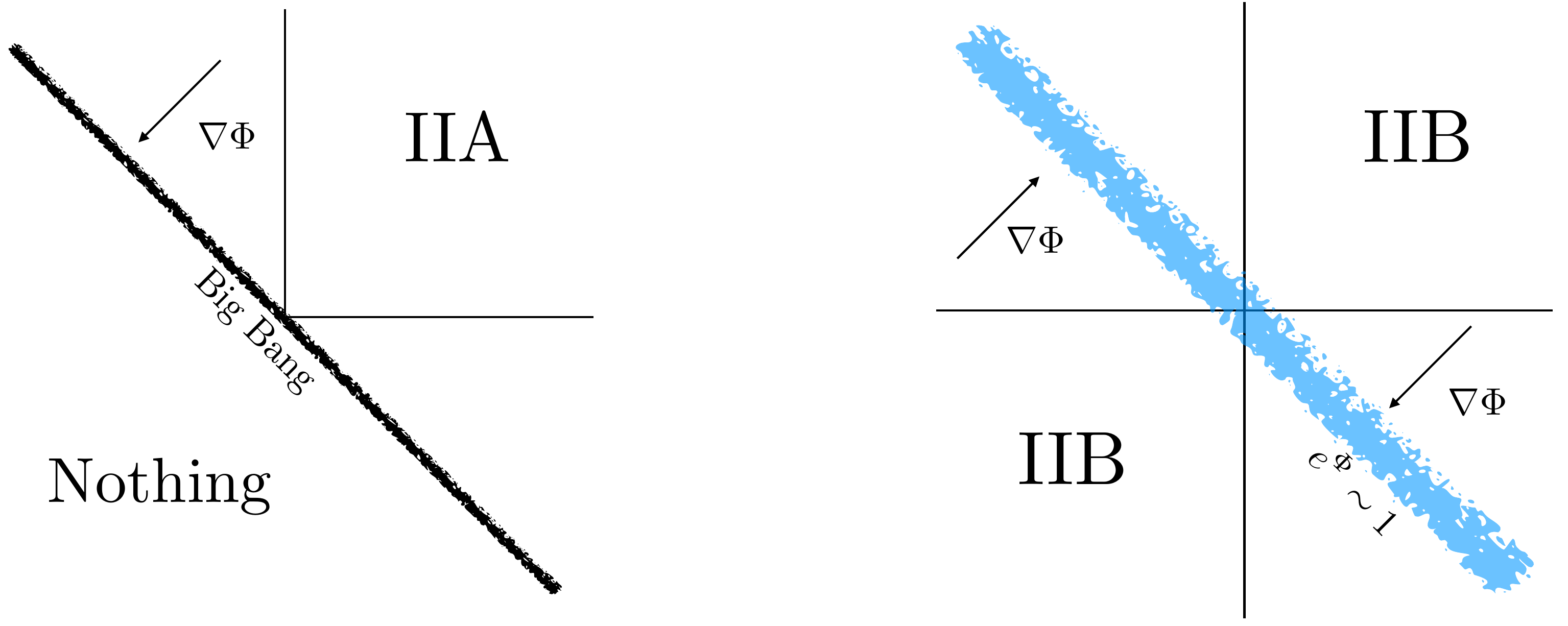}
\caption{\small The type II linear dilaton backgrounds are BPS solutions. In IIA (left panel), the strong coupling region can be uplifted to a light-like Big Bang singularity \cite{Craps:2005wd}, where the M-theory circle blows up and the scale factor vanishes. In the Figure, the thick black diagonal line depicts the Big Bang, with ``nothing'' to the bottom-left of it. This geometry is, therefore, a true end-of-the-world boundary for IIA string theory. The arrow depicts the direction of increasing dilaton. In type IIB (right panel), a possible resolution of the IIB strong coupling region: an application of S-duality to the BPS background reverses the sign of the dilaton gradient, resulting in a second copy of a IIB weakly coupled region in the bottom left of the diagram. Therefore, in this UV completion the BPS linear dilaton background would not be  a valid cobordism defect. It is conceivable that additional operators turned on in the RG flows studied in the main text turn the IIB case into something similar to IIA, but we do not know whether this is the case. }
\label{fig:linear2}
\end{center}
\end{figure}

The fate of the linear dilaton background in the IIB side is less clear. Just like for IIA, there is also an exact BPS background, since the solution only involves NS-NS sector fields. An application of S-duality would map strongly coupled region linear increasing dilaton field to a \emph{weakly} coupled region (see Figure \ref{fig:linear2}). So in IIB string theory, the BPS linear dilaton background seems to describe a solitonic NS-NS wave in a non-compact IIB spacetime, and not cobordism end-of-the-world brane. The only potential escape to this conclusion lies in the fact that, in the strongly coupled region where the string coupling takes values of order 1 (the blue fuzzy region in Figure \ref{fig:linear2}), the energy density becomes Planckian, and it may be that the solution cannot be continued past this point. Such case would correspond to the IIB lightlike linear dilaton background admitting UV completion in terms of a lightlike boundary, very much like its IIA counterpart. We view understanding the physics of the IIB linear dilaton background as an interesting future direction.

Finally, one may envision that there may exist UV completion of the strongly coupled core of our configuration which allows for transmission. In fact, in analogy with the type 0 case, our worldsheet construction naively provides consistent rules for how objects transform ``upon crossing the IIA/IIB wall'', which moreover agree with those in \cite{Heckman:2025wqd}. In a sense, the configuration seems to be ``ready'' for the possibility that a transmissible resolution of the wall exists. This may be regarded as indirect evidence for the possibility of such UV completion of the strongly coupled region, but is far from being a solid argument. We leave this as an interesting open question for the future.

\section{NS5 branes and inflow}
\label{sec:NS5s}

In earlier sections we have discussed the behavior of D-branes as they cross the cobordism domain wall between type 0A/0B or type IIA/IIB theories. In this section we address what would seem an even simpler question, namely the fate of NS5-branes as they cross the domain wall. We consider this question for the type 0 and type II cases, and both from a general macroscopic perspective based on symmetries/anomalies and from the microscopic description in terms of worldsheet interpolations introduced in the previous sections. The results provide interesting new insights into the worldvolume dynamics of the domain walls.

\subsection{General considerations based on symmetries}
\label{sec:ns5-general}

Let us start with general considerations about the behavior of NS5-branes as they cross the cobordism domain wall, based on macroscopic properties of the wall and the symmetries and anomalies of the system. This approach is agnostic regarding the microscopic description of the domain wall, very much in the spirit of \cite{Heckman:2025wqd}, although the fate of NS5-branes was not discussed there.

\subsubsection{Basic configuration and the anomaly puzzle}

The domain wall separates two 10d theories which only differ in the relative GSO projections on the Ramond groundstates. Hence, there is naively no effect on the NSNS field sector, and the natural guess is that, when an NS5-brane of one theory crosses the domain wall, it turns into just an NS5-brane of the other theory. This also naively fits with the idea that the NS5-brane is the dual of the fundamental string, which in the spirit of \cite{Heckman:2025wqd} we have implicitly considered is able to cross the wall without problems. 

In order to test this naive proposal, one can search for further consistency requirements, based on symmetries or anomalies of the configuration. In the type 0 case, since the worldvolume theories on the type 0A and 0B NS5-branes are non-chiral, there are no anomaly obstructions for the NS5-brane of one theory to turn into just the NS5-brane of the other. In fact, in section \ref{sec:ns5-supercritical} we will argue that this picture is precisely realized in the supercritical description of the type 0A/0B domain wall of section \ref{sec:type0}.

The situation is very different in the type II case, due to the intricate anomalies (and their cancellation) on the type IIA NS5-brane, in contrast with the non-chiral (hence anomaly-free) status of the type IIB NS5-brane. Let us recall that the type IIA NS5-brane hosts a 6d $(2,0)$ tensor multiplet, which is chiral and anomalous, and let us denote by $X_8$ its anomaly polynomial. The anomaly is canceled by an inflow from the 10d bulk due to a 1-loop coupling to the NSNS 2-form field \cite{Witten:1995em}
\bea 
S_{\rm inflow}=\int_{10d} B_2X_8=-\int_{10d} H_3X_7^{(0)}\, .
\label{10d-inflow}
\eea  
Here, $X_8$ is given by precisely the anomaly polynomial of the NS5-brane worldvolume theory. We are also using the descent relations $X_8=dX_7^{(0)}$, $\delta_\lambda X_7^{(0)}=\lambda dX_6^{(1)}$ (with  $\lambda$ a gauge variation parameter), and we have integrated by parts. We can now consider the general gauge variation  of (\ref{10d-inflow}) in the presence of an NS5-brane source $dH_3=\delta_4(\rm 
NS5)$, to get
\bea
\delta_\lambda S_{\rm inflow}=
-\int_{10d} \lambda \,H_3 dX_6^{(1)}=\int_{10d} \lambda\, dH_3 X_6^{(1)}=\int_{10d}\lambda\,\delta_4(\rm NS5)X_6^{(1)}=\lambda\,\int_{NS5} X_6^{(1)}\, ,\quad 
\eea 
so the anomaly from the 6d worldvolume fields on the NS5-brane is canceled by an inflow from the 10d bulk. The same arguments hold for a set of $N$ coincident branes, by simply including a multiplicative factor $N$.

In contrast, the 6d worldvolume theory of type IIB NS5-branes is non-chiral, and hence non-anomalous. This raises the question of whether it is possible that a type IIA NS5-brane crossing the domain wall can turn into just a type IIB NS5-brane, given that they are very different beasts from the anomaly viewpoint. One may be tempted to claim that the anomaly cancelation by inflow in the type IIA side removes the obstruction to such a simple crossing; after all, there is no overall 6d anomaly on either side. However, we now show that this naive picture cannot be correct.

Consider a configuration of a type IIA NS5-brane (for concreteness, along the directions 012349) crossing a domain wall located at $X_9=0$, and assume it turns into just a type IIB NS5-brane. We now take the $\R^5$ spanned by the directions 56789, transverse to the intersection of the NS5-brane and the domain wall, and regard it as a real cone over $\mathbb{S}^4$, with vertex at the said intersection. The configuration can be regarded as a compactification on $\mathbb{S}^4$, which runs along one spacetime direction of the resulting 6d theory. This kind of running solutions have been exploited extensively in the context of dynamical cobordisms \cite{Buratti:2021yia,Buratti:2021fiv,Angius:2022aeq,Angius:2023xtu,Huertas:2023syg,Angius:2023uqk,Angius:2024zjv,Angius:2024pqk,Apers:2025pon}. 

The ingredients in the $\mathbb{S}^4$ compactification are easily identified as they are inherited from the parent  flat space configuration, as follows. The $\mathbb{S}^4$ is split into two hemispheres, separated by the domain wall wrapped on the equator $\mathbb{S}^3$, so e.g. the northern hemisphere corresponds to type IIA theory, while the southern one corresponds to type IIB. There is one type IIA NS5-brane at the north pole, and one type IIB anti-NS5-brane (due to the orientation flip between the coordinate $X_9$ and the radial coordinate in the cone) in the south pole. 

This leads to the following puzzle: the type IIA NS5-brane produces an anomalous field content in the 6d theory after compactification on the $\mathbb{S}^4$, while the type IIB anti-NS5-brane gives a non-chiral content and hence does not cancel the anomaly (as would happen in a type IIA configuration with no domain wall, resulting in an $\mathbb{S}^4$ compactification with a type IIA brane-antibrane pair). Actually, we should recall that the anomaly is canceled by the inflow from the 10d type IIA bulk described above. But, in a compact space, Gauss' law requires that the inflow into the type IIA NS5-brane must be canceled by outflows from somewhere else in the $\mathbb{S}^4$. The only candidate region for this is the equator $\mathbb{S}^3$, namely the domain wall, since it corresponds to the only other boundary of type IIA region. This suggests that the domain wall must support a source of 6d anomaly, proportional to the amount of $H_3$-flux it encloses. There are several mechanisms we may propose for this, which we consider next.

\subsubsection{Solutions to the anomaly puzzle}

There are several ways out of the anomaly puzzle explained in the previous section, which we discuss in turns.

One possibility is that the domain wall supports gapless degrees of freedom, which ultimately produce the necessary anomaly in the presence of NS5-branes/$H_3$-flux. This would be an exciting scenario, as it would support the existence of a non-trivial 9d CFT on the domain wall worldvolume. However, it is far from clear it is possible to propose a 9d theory which develops the 6d anomaly of $N$ $(2,0)$ tensor multiplets upon compactification on $\mathbb{S}^3$ with $N$ units of $H_3$-flux.

A simpler possibility is to give up the assumption that a type IIA NS5-brane turns into just a type IIB NS5-brane upon crossing the wall. Let us provide a simple realization of this mechanism. Imagine the domain wall carries a $U(1)$ gauge boson on its 9d worldvolume, with lagrangian
\bea 
{\cal L}_{9d}=|f - B|^2+ k f \wedge X_7^{(0)},
\eea 
where $f$ is the $U(1)$ field strength and $X_7^{(0)}$ is the Chern-Simons 7-form associated to the anomaly polynomial $X_8$. This theory has a 10d anomaly theory which is precisely the 10d type IIA coupling (\ref{10d-inflow}), which implies it sources the necessary anomaly inflow in the presence of NS5-branes. However, it also implies that this domain wall theory cannot be compactified on $\mathbb{S}^3$ with non-trivial $H_3$-flux, because the equations of motion set $df=H_3$, so the flux integral must vanish on a manifold without sources. Hence, when the domain wall surrounds a type IIB NS5-brane, there must be an extra magnetic source on the domain wall, such as the boundary of a 6-brane stretching between the domain wall and the type IIB NS5-brane. This would amount to a Freed-Witten anomaly \cite{Freed:1999vc,Maldacena:2001xj} for the IIA/IIB domain wall.

Let us now apply this idea to the domain wall on the equator $\mathbb{S}^3$ surrounding the type IIB NS5-brane at the south pole of the $\mathbb{S}^4$: in the flat space configuration we obtain that, when the type IIA NS5-brane crosses the domain wall, it turns into a type IIB NS5-brane, but a 6-brane is created with a 6d boundary forming a corner between the NS5-brane and the domain wall\footnote{The attentive reader has probably noticed that this mechanism is analogous to the process in which a type IIA NS5-brane crosses a D8-brane and triggers the creation of a D6-brane stretched between them.}, in a Hanany-Witten kind of brane creation process \cite{Hanany:1996ie}. The part of the 6d boundary lying on the domain wall should support the chiral degrees of freedom responsible for the anomaly mentioned above.

One may regard the latter solution as a variant of the first. In a certain sense, we have described a 9d worldvolume theory which, in the presence of $H_3$-flux, leads to degrees of freedom producing the 6d anomaly. The price we have paid is that the extra degrees of freedom are not propagating in 9d, but are instead supported in a 6d sub-locus which breaks part of the 9d Poincar\'e symmetry. This is actually the simplest way to accommodate the existence of extra degrees of freedom without demanding the existence of an exotic and yet unknown (if existing at all) non-trivial gapless 9d theory.

The above scenario suggests an even simpler possibility: It may happen that the NS5-brane of one theory is unable to move across the domain wall to the other theory. In this setup, an NS5-brane along 01234 and directed towards the domain wall in the direction 9 could, instead of crossing it, be deflected in the transverse directions as an NS5-brane spanning 01234 and one semi-infinite line in the $\R^4$ within the domain wall. In this description, the piece of NS5-brane inside the domain wall plays the role of 6d defect supporting the chiral degrees of freedom. In the description as a running solution in an $\mathbb{S}^4$ compactification, there is a type IIA NS5-brane in the north pole, and no type IIB NS5-brane in the south pole, but there is a type IIA anti-NS5-brane at a point in the domain wall on the equator $\mathbb{S}^3$, so that the 6d anomalies of the brane-antibrane system cancel out.

The discussion of the realization of these mechanisms in the microscopic description of type 0A/0B and IIA/IIB domain walls will be carried out in section \ref{sec:ns5-supercritical}

\subsubsection{Crossing of $\C^2/\Z_N$ singularities}

Before finishing, we would like to mention that there is an analogous discussion of the crossing of gravitational instantons (a.k.a. KK monopoles, or $\C^2/\Z_N$ sigularities) across cobordism type 0A/0B or IIA/IIB domain walls. In fact the discussion is related to that of NS5-branes via T-duality, although this requires extra assumptions about the behavior of the domain wall under T-duality, hence we rederive the arguments directly in the $\C^2/\Z_N$ picture. 

In the type 0 case, there are no anomalies, hence no obvious obstruction to a type 0A $\C^2/\Z_N$ singularity to propagate to the type 0B side as just a $\C^2/\Z_N$ singularity. This looks natural, since it amounts to taking the flat space type 0A/0B domain wall configuration, and performing a $\Z_N$ orbifold in the directions 5678, so that the singular locus stretches along the direction 9 across the domain wall. 

On the other hand, our arguments show that things cannot be as simple in the type II case. In particular, the type IIB $\C^2/\Z_N$ singularity supports an anomalous 6d $(2,0)$ theory, associated to the localized twisted sector fields at the singularity, whose anomaly is canceled by an inflow from the 10d bulk\footnote{This may be understood from the chiral 10d fields in the untwisted sector which remain locally dynamical at the fixed locus and lead to  localized 6d anomaly. Equivalently, one may regard the $\C^2/\Z_N$ geometry as an $\mathbb{S}^1$ fibered over $\R^3$, with the fiber degenerating (with multiplicity $N$) at a point, and check that the KK modes of the chiral 10d fields in this $\mathbb{S}^1$ compactification lead to a 9d 1-loop Chern-Simons coupling $\int_{9d} A_1 X_8$, with $A_1$ the KK gauge field (see \cite{Garcia-Etxebarria:2015ota} for a derivation in the 10d and supercritical constructions). This is the T-dual of the $\mathbb{S}^1$ compactification of the type IIA coupling (\ref{10d-inflow}).}. Hence, it cannot simply turn into a type IIA $\C^2/\Z_N$ singularity (which is non-chiral and requires no inflow from the 10d bulk), but rather must involve new non-trivial ingredients at the domain wall. This can be again shown using a cone construction in the $\R^5$ spanned by 56789 (quotiented by a $\Z_N$ acting on 5678). The cone is then described as a running solution of the 6d theory obtained by compactifying the 10d configuration on $\mathbb{S}^4/\Z_N$, with the domain wall wrapped on the equator $\mathbb{S}^3$ (quotiented by a freely acting $\Z_N$), and two (differently oriented) $\C^2/\Z_N$ singularities at the type IIA and IIB poles. It is interesting to notice that this implies that it is not possible to simply take the 10d flat space domain wall configuration and orbifold it by a $\Z_N$ acting on 5678. We will later on see how our microscopic description of the type IIA/IIB domain wall provides a simple explanation of this fact.

In the following section we study the behavior of NS5-branes in their crossing of the domain wall in the explicit worldsheet interpolation description (and their possible supercritical interpretations). The behavior in both the type 0 and type II cases leads to interesting realizations of several of the above ideas. We will also consider $\C^2/\Z_N$ singularities and find satisfactory answers for all the above puzzles.

\subsection{Microscopic description of the crossing via worldsheet cobordism}
\label{sec:ns5-supercritical}

In this section we analyze the above issues in the specific description of the domain wall in terms of explicit worldsheet interpolations (and their possible supercritical interpretation). As explained in the previous section, we can equivalently choose to carry out the discussion in terms of crossing of $\C^2/\Z_N$ orbifold singularities. The discussion can be easily translated into the language of its T-dual worldsheet CFT, which describes NS5-branes. 

In general terms, the physical realization of the worldsheet interpolation corresponds to including extra 2d fermionic and bosonic fields, which are gapped at generic values of a spacetime coordinate $X_9$, but become gapless at the location of the wall, $X_9=0$, making the theory non-compact and allowing for modes to escape in the extra bosonic directions. This provides a natural resolution to the crossing of $\C^2/\Z_N$ or NS5-branes, as follows. 

The $(1,1)$ worldsheet theory is obtained by tensoring a free sector describing flat spacetime in the directions 01234, times the non-trivial CFT describing the transverse directions 5678 to the NS5-branes or $\C^2/\Z_N$, times the interpolating CFT describing the extra degrees of freedom fibered over the direction 9. Let us focus on the type II setup, in the effective intermediate description with a single additional multiplet $X$ (formerly denoted by $\Phi_8$ in section \ref{sec:typeII}). This implies that the configuration describes NS5-branes (or $\C^2/Z_N$ singularities) in a locus with two components, one given by $X=0$, describing NS5-branes stretching along $X_9$ and hence crossing the domain wall, and another given by $X_9=0$, describing NS5-branes localized onto the domain wall and escaping through the extra dimensions. Since both loci meet at $X_9=X=0$, the scenario is better described as one type IIA NS5-brane incoming from $X_9<0$, which reaches the domain wall at $X_9=0$ and bends away in the extra direction $X>0$, and simultaneously a type IIB NS5-brane incoming from $X<0$ inside the domain wall at $X_9=0$, and bending at $X=0$ into a IIB NS5-branes along $X_9>0$. Clearly, the same kind of configuration is obtained if one works with $\C^2/\Z_N$ singularities, so we continue the discussion in terms of NS5-branes.

One may object that this picture does not correspond to a proper description in the 10d critical theories, since it involves the extra multiplets, which are generically gapped and should be integrated out. Indeed, the above picture should be interpreted as an intermediate step before integrating out the extra fields. After doing that, the NS5-brane branches which span $X_9=0$, $X\neq 0$ will end up bending into one 1d line within the domain wall. This is analogous to the result in \cite{Garcia-Etxebarria:2015ota} in the supercritical context, explaining that topological defects in the extra dimensions become defects in the 10d critical theory after integrating out the extra fields. This leads to localized degrees of freedom on the domain wall, albeit in a way which breaks spontaneously the rotational symmetry in the directions 5678. This picture was in fact a solution to the anomaly puzzles presented in the previous section. We note that it is also consistent with the interpretation of the type IIA/IIB worldsheet cobordism as back-to-back bubbles of nothing, since the two kinds of NS5-branes have a consistent embedding in the two corresponding IIA/IIB regions. A type IIA NS5-brane pointed towards the domain wall scatters off along the domain wall, while the reverse process occurs for a type IIB NS5-brane on the other side. 

It should be possible to construct configurations which isolate just one of these two processes, describing the scattering off of an NS5-brane on one of the sides, with no effect on the other; these configurations cannot however be described as a simple $\Z_N$ orbifold of the flat space configuration (or its T-dual), so we will not attempt to describe them.

Let us now comment on the fact that in the type 0A/0B domain wall there is no anomaly obstruction for an NS5-brane of a theory to cross the domain wall and turn into just an NS5-brane of the other. A possible interpretation in the context of the supercritical realization of the domain wall is that the scalar potential in (\ref{type0-couplings}) stabilizes (by higher order terms) the extra bosonic field at $X=0$ even at $X_9=0$, and the above argument does not apply at face value. Hence, type 0 NS5-branes are not forced to bend as they hit the 0A/0B domain wall. A similar conclusion holds for $\C^2/\Z_N$ singularities across the cobordism 0A/0B domain wall.

It is satisfying that the description of domain walls in terms of worldsheet cobordisms provides a microscopic understanding of the behavior of exotic objects like NS5-branes and $\C^2/\Z_N$ singularities across type 0A/0B and IIA/IIB domain walls. It would be interesting to extract additional information on the worldvolume dynamics of the domain walls by using these and other objects crossing it. We leave these interesting explorations for future work.

\section{Conclusions: A Worldsheet Odyssey}
\label{sec:conclusions}

In the last few years our understanding of non-supersymmetric branes has significantly expanded, spearheaded by the cobordism conjecture. In spite of this progress, our understanding of many cobordism defects and/or their worldvolume dynamics remains superficial or limited to quantities protected by symmetries. In this paper we have tried to apply worldsheet techniques to better understand some of the more interesting bordism defects -- domain walls between 0A/0B or IIA/IIB string theories.

After reviewing the obstructions to the existence of a worldsheet description for these objects, we tried to evade them by using non-compact bordisms. In some sense, we can say we succeeded: we found non-compact worldsheet QFT's that interpolate between the 0A and 0B theories, and also between IIA and IIB. To do the latter, we had to study a (1,1) version of the Gross-Neveu model which, to our knowledge, has not appeared in the literature so far. In any case, the actual domain walls obtained from following the RG flow to its IR limit all have strongly coupled regions, meaning that a full understanding of these defects lies beyond the reach of perturbative string theory, as was expected from anomaly considerations.

Furthermore, the strongly coupled descriptions we found for the vicinity of 0A/0B or IIA/IIB domain walls are old acquaintances: On each side of the domain wall, they correspond to linear dilaton backgrounds, a well-studied solution of critical 10d string theory \cite{Craps:2005wd}. In  these solutions, which depend on a single null coordinate, the dilaton ranges from very small to very large values, where perturbative string theory breaks down. To understand the domain walls we have constructed, one needs to understand this strongly coupled region.  

 As explained in \cite{Hellerman:2006ff} there is more than one UV completion to the linear dilaton background, so there are different possibilities. In the type IIA side a complete UV description of a BPS version of this background was provided in \cite{Craps:2005wd} in terms of a lightlike end-of-the-world boundary. This may extend to a similar behavior for the IIB version of the lightlike linear dilaton background, but there may be other possibilities, such as an NSNS wave separating two connected weakly coupled IIB regions, related by S-duality. If our linear dilaton backgrounds uplift to something close to their BPS counterparts, then the conclusion is that we constructed an end-of-the-world brane for IIA, but failed for IIB. Since the linear dilaton physics is UV sensitive, however, what is the ultimate fate of our construction in the IIB side remains to be seen. One may in fact envision that there may exist a resolution of the strongly coupled region which allows for transmission, and it is perhaps suggestive that the worldsheet theory seems to be ready for this, given that it provides consistent rules for the transformation of different objects as they cross the wall. In any event, addressing the resolution (or possible resolutions) of lightlike linear dilaton backgrounds in critical string theories seems a most important open question for future work. In this, the non-perturbative approach of \cite{Torres:2026vxx} may be helpful.

We note that for any cobordism class represented by a critical string worldsheet CFT $X$, we can tensor it with the linear dilaton background, and if the strong coupling region can be interpreted as a boundary, we have attained a cobordism defect for the cobordism class represented by $X$. Thus, if the linear dilaton counts as a boundary, we can construct a cobordism defect for any background in string perturbation theory! Just tensor any worldsheet with the linear dilaton background. For instance, doing this for the AOB background \cite{Aharony:2007du} results in a worldsheet description of an R7 brane. This R7 brane solves the spacetime equations of motion, but it is expanding at the speed of light due to the light-like linear dilaton and is, therefore, not stable, at least in this configuration; the arguments in \cite{Heckman:2025wqd} may not apply, since they hold for EFT instabilities only, while we also have a string-sized linear dilaton gradient. On the other hand, see \cite{Cavusoglu:2026xiv} for a recent analysis providing a suitable backreacted profile for the R7 far away from its core region. In these solutions, the asymptotic profile for the R7 has a dipole moment, thus leading to a stronger breaking of Lorentz symmetry than the bare minimum expected for a codimension-1 object. The linear dilaton R7 branes we would get from the techniques in this paper share this feature, as well as the constructions in \cite{Torres:2026vxx}. Perhaps this is all telling us that low-codimension objects must necessarily break Lorentz symmetry explicitly, as suggested in \cite{Blumenhagen:2022mqw}. 

We have also discussed the interesting configurations of NS5-branes (and $\C^2/\Z_N$ singularities) crossing the domain walls, in particular in the IIA/IIB case, and argued they reveal key information about their worldvolume dynamics. Our model independent analysis reveals that either there is a non-trivial gapless 9d CFT describing the domain wall worldvolume, or else it carries a topological sector reproducing a Freed-Witten anomaly associated to worldvolume 6d defects, ensuring that IIA NS5-branes directed towards the domain wall are either deflected along it, or ``peeled off'' of their anomaly to be allowed to cross as IIB NS5-branes (dressed with additional suspended 6-branes). We have further exploited our explicit worldsheet description of the domain wall to provide concrete arguments favoring the presence of such topological sector. A systematic clarification of the possible worldvolume topological couplings and their dynamical realization in our setup in a promising interesting direction.

When we set out, our goal was to find a way to construct cobordism defects in perturbative string theory. If the linear dilaton backgrounds we found can indeed be used as cobordism end-of-the-world branes for all perturbative string theories then, in some sense, our quest has succeeded -- we would have a simple proof of the Cobordism Conjecture for an ample class of backgrounds. But much like Odysseus' homecoming, the success is bittersweet -- the worldsheet description is not really useful to understand novel phenomena like the ones proposed in \cite{Yonekura:2024spl}, or the detailed dynamics of cobordism defects. Furthermore, the worry that in some cases (like in IIB), the linear dilaton backgrounds are not end-of-the-world defects still looms. We hope to come back to these fascinating questions soon in a novel, (target) Space Odyssey!

\section*{Acknowledgments}
 We thank Chiara Altavista, Roberta Angius, Jos\'e Calder\'on-Infante, Matilda Delgado, Markus Dierigl, Simeon Hellerman, Luis Ib\'a\~nez, and Ethan Torres for useful discussions. We thank the Spanish Research Agency (Agencia Estatal de Investigacion) through the grants IFT Centro de Excelencia Severo
Ochoa CEX2020-001007-S, PID2021-123017NB-I00 and PID2024-156043NB-I00, funded by MCIN/AEI/10.13039/501100011033 and by ERDF A way of making Europe. The work by E. A. is supported by the fellowship LCF/BQ/DI24/12070005 from ``La Caixa'' Foundation (ID 100010434). MM is currently supported by the RyC grant RYC2022-037545-I and project EUR2024-153547 from the AEI. C. W. is supported by program PIPF-2024/TEC-34293 from Comunidad de Madrid.

\newpage

\appendix

\section{Supercritical strings, closed tachyon condensation and all that}
\label{app:supercritical}

In this Appendix we provide some useful background regarding supercritical string theories of different kinds. Although these theories have a timelike linear dilaton, so they always contain a strongly coupled region in the far past (see \cite{Angius:2022mgh} for a proposed resolution in the bosonic theory), they are otherwise weakly coupled and well behaved at later times, and they provide an interesting physical interpretation for the constructions of worldsheet interpolations in the main text. We mainly follow \cite{Hellerman:2006ff}, to which we refer the reader for further details.

\subsection{Supercritical type 0}
\label{app:type0}

The worldsheet theory of supercritical type 0 theory in flat $D=10+n$ dimensions are described (ignoring ghosts) by $10+n$ 2d real fields $(X^M, \psi^M)$ forming $(1,1)$ multiplets. The cancelation of the central charge requires including a timelike linear dilaton background $V^M$, satisfying
\begin{equation}
V^M V_M=-\frac{n}{4\alpha'}.
\label{dilaton-back}
\end{equation} 
The theory has a gauged $\mathbb{Z}_2$ symmetry generated by $(-1)^{F}$, where $F$ is the overall (left- plus right-moving) fermion number, leading to a diagonal modular invariant partition function. This corresponds to a GSO projection implemented by summing over spin structures common to both left- and right-moving sectors. For even dimensions there is a two-fold choice, leading to the supercritical 0A and 0B theories, associated to choosing opposite or equal GSO projections on the left- and right-moving Ramond groundstates (equivalently, flipping the chirality of the spinor groundstate in e.g. the left-moving Ramond sector). The light spectrum in the NSNS sector contains a real closed tachyon scalar, and massless graviton, 2-form and dilaton fields, and the RR sector contains two sets of massless $p$-form potentials (with $p$ odd/even for the 0A/0B theories).

We will be interested also in the odd-dimensional case. In this case, there is no notion of chirality for the Ramond groundstates, so there is only one possible theory. The NSNS sector is as above, but the RR sector contains a single copy of massless $p$-form potentials for any even and odd degrees. We will discuss this in more detail later on.

There is interesting physics associated to the closed string tachyon in the NSNS sector in these theories, as we now review. Following \cite{Hellerman:2006ff}, the tachyon couples as a worldsheet $(1,1)$ superpotential, 
\begin{equation}
\Delta {\cal L}\, =\, \frac{i}{2\pi}\int d\theta^+d\theta^-{\cal T}(X)\, =\; -\frac{1}{2\pi} \sqrt{\frac{\alpha'}{2}} F^M \partial_M {\cal T}(X) + 
\frac{i\alpha'}{4\pi} \partial_M\partial_N{\cal T}(X)\, {\tilde \psi}^M \psi^N\, ,
\label{wsupo}
\end{equation}
where tilded and untilded fields correspond to left- and right-moving fields, and $F^M$ is the auxiliary field for the superfield $X^M$. Using their equations of motion $F^M\sim G^{MN}\partial_N {\cal T}$, we get a worldsheet potential term
\begin{equation}
V\, =\, \frac{\alpha'}{16\pi} G^{MN}\partial_M{\cal T}\partial_N{\cal T}\, .
\label{pot}
\end{equation}

One can now study tachyon profiles and describe the corresponding spacetime interpretation. In particular we focus on the lightlike tachyon profiles studied in \cite{Hellerman:2006ff}, which were shown to lead to CFT's solvable in an $\alpha'$-exact way (1-loop exact in $\alpha'$, to be more precise), and to reduce the number of spacetime dimensions of the type 0 theory (a process dubbed ``dimension quenching''). For instance, we can remove one dimension $X$ by considering
\bea
{\cal T}=\mu \exp(\beta X^+) X^2\, 
\label{tachyon-quench1}
\eea
where $\mu$ is an arbitrary coefficient, and $\beta$ is fixed so that the tachyon deformation is an exactly marginal operator of the CFT.
In the following we will often skip the dependence on the lightlike coordinate $X^+$ for simplicity, but it should be implicitly kept in mind. Using (\ref{wsupo}) and (\ref{pot}), at late $X^+$ this tachyon profile corresponds to a mass term for the 2d boson $X$ and its fermion partner $\psi$, namely
\bea
\Delta {\cal L}\, \sim \mu^2  X^2+\mu {\tilde\psi}\psi
\eea
Integrating out these massive 2d fields leads to the worldsheet content of the theory in $(D-1)$ dimensions. The integrating out process renormalizes the spacetime background such that the linear dilaton profile adjusts and the central charge continues to cancel  \cite{Hellerman:2006ff}. As we discuss in section \ref{sec:type0-worldsheet-interpolation}, starting with odd dimension $D$, the sign of the mass term $\mu$ determines if the resulting theory in an even number $(D-1)$ of spacetime dimensions is the type 0A or 0B theory.

\subsection{Supercritical heterotic}
\label{app:heterotic}

We now move to the realization of similar ideas for heterotic strings. There exist several supercritical versions of heterotic strings, considered e.g. in \cite{Hellerman:2004zm,Hellerman:2006ff}. We start with the simplest diagonal modular invariant theory, denoted as HO$^{+/}$ in those references. The $(0,1)$ worldsheet content for this theory in $D=10+n$ flat dimensions is (ignoring ghosts) given by $10+n$ bosons $X^M$, $10+n$ right-moving fermions $\psi^M$ and $32+n$ left-moving current algebra fermions ${\tilde\lambda}^a$.

The partition function is the diagonal one, namely the only $\Z_2$ gauge symmetry is by $(-1)^F$, where $F$ is the overall left- plus right-moving worldsheet fermion number. The theory contains the graviton, 2-form and dilaton fields, as well as $SO(32+n)$ gauge bosons propagating in $10+n$ dimensions, and there is a set of tachyons ${\cal T}^a$ transforming in the vector representation of $SO(32+n)$. On the other hand, the Ramond groundstates are massive, so these theories have no massless fermions.

These theories are related by dimension quenching via closed tachyon condensation. The spacetime tachyon profile ${\cal T}^a(X)$ couples on the worldsheet as a $(0,1)$ superpotential
\bea
\Delta {\cal L}=-\frac 1{2\pi}\int d\theta_+ \sum_a {\tilde \lambda}^a {\cal T}^a(X)
\eea 
By expanding in components the right-moving superfields, and integrating out auxiliary fields, we obtain a scalar potential and fermion couplings
\bea 
\Delta {\cal L}=-\frac 1{8\pi}\sum_a ({\cal T}^a(X))^2+\frac{i}{2\pi}\sqrt{\frac{\alpha'}2}\partial_M {\cal T}^a(X) {\tilde \lambda}^a\psi^M
\label{heterotic-tachyon-condensation}
\eea 
As discussed in \cite{Hellerman:2006ff}, there are lightlike tachyon condensation processes reducing the number of dimensions, which can be studied in an $\alpha'$-exact manner. Let us consider the particular case of removing $n$ supercritical dimensions to end up in a 10d theory. The corresponding tachyon profiles are of the form
\bea
{\cal T}^a(X)=\mu\sqrt{\frac 2{\alpha'}}e^{\beta X^+} X^{9+a}\, ,
\eea 
with $\beta$ again determined by the requirement that this is a marginal deformation. The above can be generalized to an arbitrary linear dependence of the tachyon with the coordinates, and for simplicity we have assumed it to the proportional to the identity.

The resulting terms in (\ref{heterotic-tachyon-condensation}) become
\bea 
\Delta{\cal L}=-\frac{\mu^2}{4\pi}e^{2\beta X^+} \sum_{a=1}^n (X^{9+a})^2+\frac {i\mu}{2\pi}e^{\beta X^+}{\tilde\lambda}^a\,(\,\psi^{9+a}+\beta X^{9+a}\psi^+\,)
\eea 
At late $X^+$ the tachyon condensate removes the $n$ supercritical coordinates and reduces the extra fermions so that the gauge symmetry becomes just $SO(32)$. As in previous examples, the one-loop corrections to the background are 1-loop exact in the $\alpha'$-expansion, and they modify the metric and dilaton such that the central charge adjusts to the new field content. In particular, since the endpoint is a critical 10d theory, the final linear dilaton background is lightlike. Even though the resulting critical 10d theory has gauge group $SO(32)$, it corresponds to the non-supersymmetric heterotic because its partition function is the diagonal modular invariant, in other words, there is no gauged $\mathbb{Z}_2$ imposing the supersymmetric GSO projection in the right-moving sector.

It is however easy to modify these theories to include massless fermions and provide supercritical heterotic theories which relate to the 10d supersymmetric $SO(32)$ heterotic via closed tachyon condensation \cite{Hellerman:2006ff}, as we now review. The procedure is to orbifold the above diagonal theory by a further $\mathbb{Z}_2$ generated by $(-1)^{f}\cal R$, where 
${\cal R}:X^{9+a}\to -X^{9+a}$ for $a=1,\ldots,n$
(namely, flips the $n$ supercritical coordinates), and $f$ flips the sign of all the $32+n$ ${\tilde \lambda}^a$ and the extra right-moving fermions $\psi^{9+a}$, $a=1,\ldots,n$. The product of this generator with $(-1)^F$, with $F$ the overall worldsheet fermion number, leads to an operator implementing the GSO projection on 10 of the right-moving fermions, just like in the 10d supersymmetric heterotic theory.

The spectrum of the theory contains the untwisted sector, which at the massless level contains the $D$-dimensional fields in the NS sector of the parent theory, subject to invariance under the geometric reflection ${\cal R}$; the tachyon is intrinsically odd under it, so it is forced to vanish at the fixed locus of ${\cal R}$, which is the 10d critical slice. There is also a twisted sector, which produces a gravitino and dilatino, which are jointly described as a vector-spinor of the $SO(1,9)$ Poincar\'e symmetry at the fixed locus; it also contains 10d spinors in the adjoint of $SO(32)$.

This orbifold is compatible with tachyon condensation, since the superpotential pairs up coordinates and left-moving fermions which are odd under the orbifold action, so it is invariant. At late $X^+$, the tachyon profile removes the $n$ supercritical dimensions, so the endpoint is a critical 10d theory with a lightlike linear dilaton background. In the absence of the extra coordinates, the generator of the orbifold in the critical 10d slice is just $(-1)^{F_L^{ws}}$, which flips the sign of the 32 left-moving current algebra fermions. Its product with $(-1)^F$ is  $(-1)^{F_R^{ws}}$, with $F_R^{ws}$ being the right-moving worldsheet fermion number, implementing the supersymmetric GSO projection. Hence, the resulting theory is the 10d supersymmetric $SO(32)$ heterotic theory (with a non-trivial lightlike linear dilaton background). Indeed, the resulting spectrum can also be recovered from the supercritical perspective. The untwisted spectrum contains the 10d graviton, 2-form and dilaton, and $SO(32)$ gauge bosons (and no tachyon because it was forced to vanish on the critical 10d slice), while the twisted sector reduces to the 10d gravitino and dilatino, and the $SO(32)$ gauginos.

\subsection{Supercritical type II}
\label{app:typeII}

We now overview two possible supercritical extensions of type II string theories, constructed in \cite{Hellerman:2004qa} and \cite{Hellerman:2004zm}. In either, the starting point is the supercritical type 0 theory, namely the worldsheet content (ignoring ghosts) is given by $10+n$ bosons $X^M$ and fermions $\psi^M$, and a $\Z_2$ gauge symmetry of overall (left plus right) worldsheet fermion number. 

In order to obtain a type II theory we must gauge an additional $\Z_2$ acting as e.g. left-moving worldsheet fermion number. It is well-known that this extra $\Z_2$ symmetry has an anomaly, which cancels when the number of extra dimensions is multiple of 8. Hence, there are supercritical versions of type IIA and type IIB string theories for dimensions $D=2+8k$. Actually, there are two different series of such theories, denoted in \cite{Hellerman:2004zm} as the standard series, for $D=10+16k$ (with partition function being a direct generalization of that of 10d type II theories) and the Seiberg series, for $D=2+16k$ (which are a generalization of the 2d type II theories in \cite{Seiberg:2005bx}). The fact that the structure of the partition function has a mod 16 periodicity is a reflection of the backfiring bosonization phenomenon \cite{BoyleSmith:2024qgx} (see also \cite{Heckman:2025wqd}), a subtle distinction between gauging the fermion parity symmetry (captured by a $\Z_{8}$ anomaly) and summing over spin structures (captured by a $\Z_{16}$ anomaly). We refer the readers to \cite{Hellerman:2004zm} for the detailed partition functions, and simply mention that the resulting theories are free of tachyons, and there is no known worldsheet physical process connecting theories in different dimensions.

There is a second possibility to define supercritical versions of type II, which is valid for any even number of supercritical dimensions, $n=2k$ \cite{Hellerman:2004qa}. Let us denote the supercritical dimensions by $X^i$, $Y^i$, $i=1,\ldots,k$. The type II (A or B) theories are obtained from the $(10+2k)$-dimensional type 0 (A or B) theories by gauging a $\Z_2$ which acts as $(-1)^{f}R$, where $R$ is a geometric action $Y^i\to -Y^i$ and
$(-1)^{f}$ anticommutes with the left-moving supercharge (namely it flips left-moving fermion partners of $X^i$, leaving left-moving fermions parterns of $Y^i$ invariant, and viceversa for their right-moving fermion partners). The resulting theories contain an untwisted sector propagating in $10+2k$ dimensions, containing a tachyon, and the massless graviton, 2-form and dilaton. There is a twisted sector localized on the $(10+k)$-dimensional subspace $Y^i=0$, containing gravitinos and dilatinos obtained as a the product of a vector of $SO(1,9+k)$ on the fixed locus times a bi-spinor of the $SO(1,9+k)\times SO(k)$ symmetry of the full spacetime. 

 The tachyon lives in the bulk spacetime, which is locally a type 0 theory, so it couples to the worldsheet as a $(1,1)$ superpotential, c.f. (\ref{wsupo}). This class of theories admits tachyon condensation processes, which can be studied in an $\alpha'$-exact way via lightlike tachyon profiles of the form
 \bea
 W\sim \int d^2\theta \, e^{\beta X^+} \sum_{i=1}^k \mu_i X^i Y^i\, ,
 \label{supo-supercritical-typeii}
 \eea
 with $\beta$ fixed to make the above term marginal, and we have chosen a diagonal mass matrix with coefficients $\mu_i$. This describes the removal of extra dimensions in pairs, and the 1-loop exact integration out of the massive degrees of freedom redefines the metric and dilaton background, so that the central charge readjusts. When the $n=2k$ dimensions are removed, the endpoint is the critical 10d type IIA/B string theory in the presence of a lightlike linear dilaton background.

 The above two procedures to define supercritical versions of type II theories can be combined in dimensions $D=2+8k+2k'$ \cite{Hellerman:2004zm}. We consider the type 0 theory, and split the bosonic coordinates in a set of $2+8k+k'$ denoted $X^M$, and $k'$ denoted by $Y^i$. We now quotient the corresponding $D$-dimensional type 0 theory by a $\Z_2$ acting as $(-1)^{f}R$, where $R$ is a geometric action acting as $Y^i\to -Y^i$ on these $k'$ supercritical coordinates, and $(-1)^{f}$ anticommutes with the left-moving supercharge. These theories have bulk tachyons in their untwisted sector if $k'\neq 0$, and their condensation removes coordinates of $X$ and $Y$ types in pairs. The endpoints of these tachyon condensation processes occur when the $2k'$ dimensions are removed, and lead to the type II theories in $2+8k$ dimensions of the standard or Seiberg series, which are tachyon-free.

\section{Interpolating between different spin structures on \texorpdfstring{$\mathbb{S}^1$}{S1}}
\label{app:spin-structures}

In this Appendix we describe another example of non-compact bordisms in worldsheet theories, providing a paradigmatic model of the construction. We aim to construct a family of 2d $(1,1)$ theories describing an interpolation between compactifications on $\mathbb{S}^1$ with different spin structures in the target spacetime. The $\mathbb{S}^1$ compactification with periodic or antiperiodic fermion boundary conditions are in different classes of spin bordism, hence the worldsheet sigma models describing such configurations in target space cannot be connected by  any continuous interpolation of compact worldsheet theories. We will now show how this can be done if we allow the interpolation to pass through a non-compact 2d QFT.

Consider a 2d $(1,1)$ theory describing a target $\R\times\mathbb{S}^1$, parametrized by a non-compact multiplet $Y$ and a compact one $X\sim X+2\pi R$, with canonical kinetic terms, and take the spin structure on the $\mathbb{S}^1$ to be periodic. In order to describe an $\mathbb{S}^1$ compactification, we can introduce a superpotential mass term lifting the non-compact direction $Y$, namely
\bea
W\sim Y^2\, .
\eea
We can generalize this idea and connect the theory with an $\mathbb{S}^1$ compactification with antiperiodic fermions in the target spacetime. Consider the complex coordinate
\bea
Z=e^{(Y+iX)/R}\, .
\eea 
In the $Z$-plane, the  $\R\times\mathbb{S}^1$ cylinder has been flattened out, so that the $\mathbb{S}^1$'s at fixed $Y$ correspond to circles of constant $|Z|$, the asymptotic circle at $Y\to -\infty$ corresponds to the origin, and that at $Y\to\infty$ corresponds to the circle at infinity in the $Z$-plane. Hence, circles in the $Z$-plane which surround the origin are wrapped non-trivially around the $\mathbb{S}^1\subset \R\times\mathbb{S}^1$ and have periodic spin structure, while circles which do not surround it are trivial in $\R\times\mathbb{S}^1$ and must have antiperiodic spin structure.

\begin{figure}[htb]
\begin{center}
\includegraphics[scale=.4]{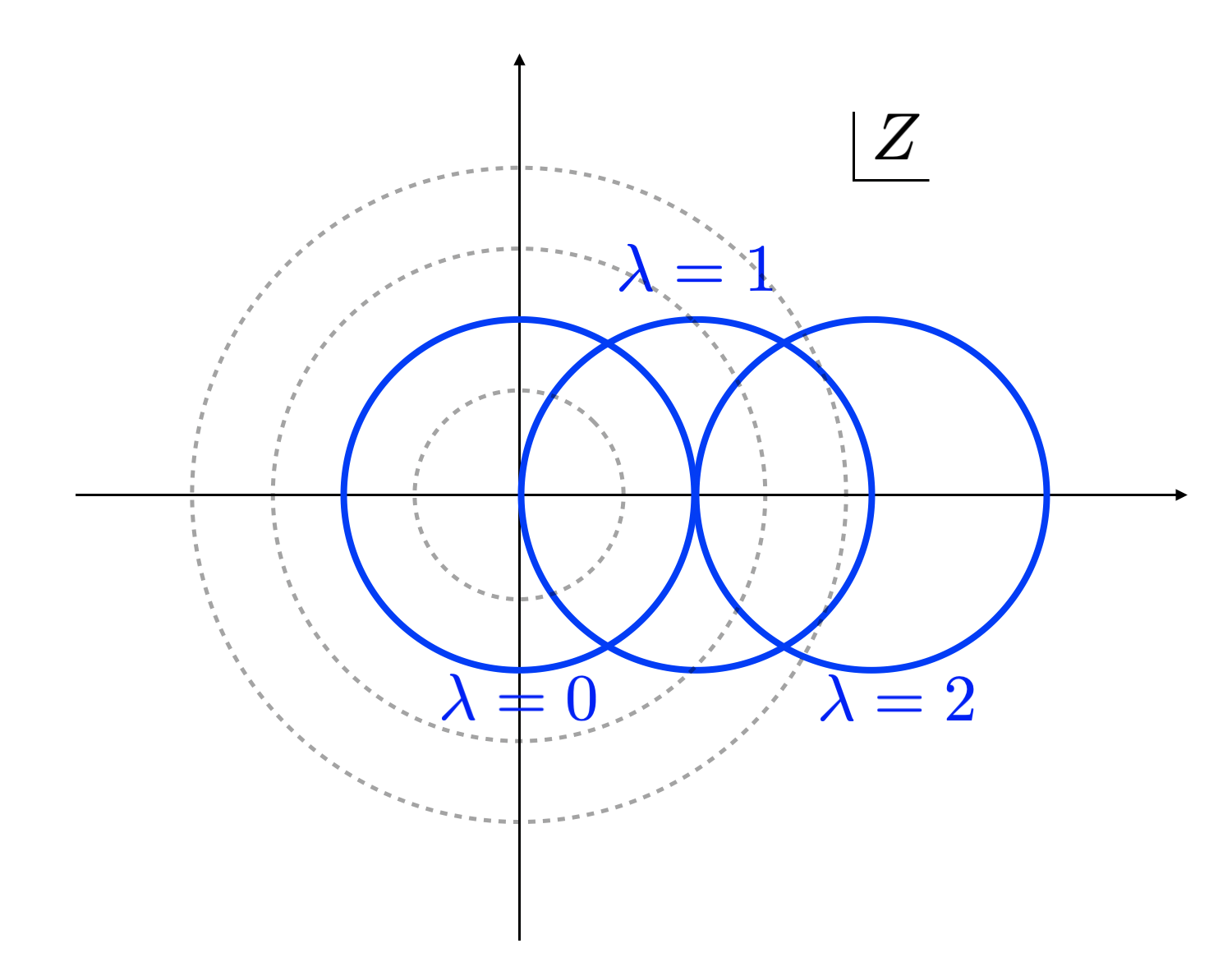}
\caption{\small Structure of $\R\times\mathbb{S}^1$ flattened onto the $Z$-plane and its different kinds of circles. The gray dashed lines correspond to  $\mathbb{S}^1$'s at constant $Y$ in the underlying $\R\times\mathbb{S}^1$. The origin corresponds to $Y\to -\infty$ and the asymptotic circle corresponds to $Y\to\infty$. The blue circles describe a family of $\mathbb{S}^1$'s interpolating between one which surrounds the origin (labeled with $\lambda=0$) and another which does not (labeled with $\lambda=2$). The interpolation involves a configuration (labeled $\lambda=1$) for which the circle passes through the origin, and is hence non-compact.}
\label{fig:circles}
\end{center}
\end{figure}

We now consider a family of superpotential terms which restrict the dynamics of $\R\times\mathbb{S}^1$ to an $\mathbb{S}^1$ which interpolates between one which surrounds the origin and one which does not. Consider the point $Z=\lambda R$, and denote by $\Delta_\lambda(Z)$ the distance (with flat metric in the $Z$-plane, and in units of $R$) between a point $Z$ to the point $Z=\lambda R$. We then consider the family of superpotentials
\bea
W_\lambda=(\Delta_\lambda(Z)^2-1)^2=
\left[\left(e^{\frac{Y}{R}}\cos\frac{X}{R}-\lambda R\right)^2+e^{\frac{2Y}{R}}\sin^2\frac{X}{R}-1\right]^2
\eea 
with $\lambda$ varying in $[0,2]$. This superpotential restricts the dynamics to the circle $\Delta_\lambda(Z)=1$, see Figure \ref{fig:circles}. This interpolates between the configuration for $\lambda=0$, describing a circle around the origin in the $Z$-plane (non-trivial in the original $\R\times \mathbb{S}^1$, hence with periodic spin structure) and the configuration with $\lambda=2$, describing a circle not surrounding the origin (trivial in $\R\times \mathbb{S}^1$, hence with antiperiodic spin structure). The configurations cross a singular one, for $\lambda=1$, in which the circle passes through the origin, corresponding to a non-compact circle in the original $X$, $Y$ coordinates. 

%%%%%%%%%%%
%%%%%%%%%%%

The intuition in the target spacetime is that starting from a circle with periodic spin structure, one can make it non-compact, and then make it compact again, but now with antiperiodic spin structure. Clearly, passing through the non-compact configuration allows to get rid of the non-trivial cobordism class of the periodic spin structure. This is a general lesson, valid for general compactifications, as discussed in the main text.

Let us now discuss how to connect this example with the divergent kinetic terms describing the infinite distance point at strong coupling. For each finite $\lambda$, the vacuum manifold is parametrized by some circle with coordinate ${\tilde X}$ of varying radius $R(\lambda)$. As $\lambda\to 1$, $R$ diverges, and then it becomes finite again. We can always rescale to a circle of periodicity $2\pi$ at the expense of reabsorbing the radius into a kinetic term of the form $R(\lambda)^2(\partial {\tilde X})^2$. For $\lambda\sim 1$, the length of the circle scales with $\lambda$ as
\bea
R(\lambda)\sim \frac 1{|\lambda-1|}\, .
\label{kin-lambda}
\eea 
This follows because the divergence of the radius is controlled by the divergence of the length of the small interval  $-\epsilon<{\rm Im}\, Z<\epsilon$ (at ${\rm Re}\, Z=\lambda-1$), computed with the metric in $Z$ induced from the flat metric in $(X,Y)$, via the exponential relation.

The above interpolation may be turned into a domain wall by considering a configuration in which the parameter $\lambda$ varies along one of the spacetime directions, which at leading order may be represented by $\lambda-1\sim X_9$. As usual, this should be regarded as the leading approximation to a generally more involved global behavior, for instance $\lambda = 1+\frac 2\pi\arctan X_9$ (which takes values in  $[0,2]$). In principle it should be possible to consider the region $X_9\sim 0$ (namely $\lambda\sim 1$), and dress the superpotential with a lightlike dependence in $X^+$ to turn the background into an on-shell solution, but we refrain from doing so. We simply note that the kinetic term (\ref{kin-lambda}) becomes
\bea
\frac{(\partial {\tilde X})^2}{|X_9|^2}\, ,
\eea 
 which is the same kind of divergence encountered in other non-compact interpolations in the main text, c.f. section \ref{sec:quantum}.

\section{Effect on RR forms}
\label{app:rrforms}

In this Appendix we describe the effect of the type 0A/0B domain wall on the spacetime massless fields, as one crosses from the 0A to the 0B sectors. Since the NSNS sectors are the same in both theories, we restrict the discussion to the RR fields. Our analysis provides the type 0 avatar of the phenomenon proposed in \cite{Heckman:2025wqd} for the type II case, deriving it from an 11d supercritical type 0 string interpretation. The main result, as in \cite{Heckman:2025wqd}, is that the RR potentials in one theory turn into RR fluxes in the other. This behavior is precisely compatible with the transformation of D-branes discussed in the section \ref{sec:d-branes}. It also provides a natural interpretation of the $\Z_2$ symmetry noticed in  \cite{Heckman:2025wqd} as a parity symmetry of the 11d theory.

Let us start by recalling that in 11d type 0 theory we have a RR spectrum of $p$-forms with all possible degrees. We split them according to their degree being even or odd, namely we introduce the formal sum of RR fields (to be later interpreted as potentials or field strengths)
\bea 
F_e=F_0+F_2+\ldots+F_{10}\quad ,\quad F_o=F_1+F_3+\ldots+F_9\, .
\eea 
In principle one should now propose an 11d action, describing the interaction of these degrees of freedom with the 11d tachyon, such that after the tachyon condensation removes the 11th dimension we recover half of the degrees of freedom, corresponding to the RR fields of type 0A or 0B theories. This derivation is however intractable, since it would require a knowledge of the coupling of RR fields to the tachyon, possibly to arbitrarily high powers of the tachyon field and its derivatives. In fact, it is not even clear that such a description within spacetime low-energy theory is possible, because in tachyon condensation processes the dynamics may not admit a purely effective field theory interpretation, as is familiar in open tachyon condensation (see e.g. \cite{Sen:1999mg} for a review).

Instead, we aim at constructing a 11d theory describing the physics of the RR fields {\em after} the tachyon condensation. In particular, since the tachyon condensation leads to purely 10d theory, with no propagating degrees of freedom in the 11d bulk, we are looking for a topological 11d action, with local dynamics arising only in the 10d locus where the tachyon vanishes. We will see that such a description suffices to capture the relevant physics, presumably because its topological nature is protected with respect to the dramatic phenomena of closed tachyon condensation.

The construction of the 11d topological action with local dynamics in a 10d slice is formally analogous to results in the literature about finding democratic formulations for (possibly chiral) $p$-form fields in a $d$-dimensional spacetime $\mathcal{X}_{d}$. Such construction involves auxiliary topological actions in a $(d+1)$-dimensional spacetime $\mathcal{Y}_{d+1}$ with $\mathcal{X}_{d}$ as its boundary $\partial \mathcal{Y}_{d+1}=\mathcal{X}_{d}$, at which suitable boundary conditions provide the dynamics, see e.g. \cite{Evnin:2023ypu}. In our tachyon condensation context, the 10d is not a boundary of the 11d spacetime, but a sublocus supporting the surviving local dynamical fields. However, the analogy is maintained if we regard our system as a doubling-trick version of the boundary conditions of an 11d theory defined on half-space, with the 10d slice as its boundary. In this description, the bulk topological actions and boundary terms in \cite{Evnin:2023ypu} provide a template for the bulk topological action and localized dynamical terms in our tachyon condensation system. 

Inspired by the above reference, we introduce a topological action
\bea
S_{11}=\int_{11d} (F_e\,dF_e+F_o\,dF_o)\, .
\label{action-11d}
\eea
In this and similar expressions, we should as usual expand the formal sums and keep the terms with correct degree for integration. 

We must now introduce the dynamics on the 10d slice after tachyon condensation. As explained, this is obtained by thinking about the 10d slice as the boundary of an 11d manifold, and using the doubling trick. This implies that the action on the 10d slice, i.e. the boundary conditions of the 11d theory are those in \cite{Evnin:2023ypu}, but restricted to $\Z_2$-invariant fields. In order to make this $\Z_2$ action manifest, we decompose the fields as
\bea
F_e=K_e+v\wedge {\tilde L}_o\quad ,\quad F_o=L_o+v\wedge {\tilde K}_e\, ,
\label{decomp-11d}
\eea 
where $v$ is a 1-form along the 11th coordinate (the Poincar\'e dual 1-form of the 10d slice), which is hence odd under the $\Z_2$. The boundary conditions in the 0A side correspond to a $\Z_2$ symmetry $F_e\to F_e$, $F_o\to -F_o$, hence the surviving components are $F_e=K_e$, $F_o={\tilde K}_e$. In the 0B side, they correspond to the $\Z_2$ symmetry $F_e\to -F_e$, $F_o\to F_o$, so the surviving components are $F_e={\tilde L}_o$, $F_o=L_o$.
The boundary conditions are the $\Z_2$-invariant components of the boundary terms in \cite{Evnin:2023ypu}
\bea 
-\frac 12 \int_{10d} (|F_e|^2+|F_o|^2)\quad & \Longrightarrow & \quad {\rm 0A:}\; -\frac 12\int_{10d} (|K_e|^2+|{\tilde K}_e|^2)\\
& \Longrightarrow & \quad {\rm 0B:}\; -\frac 12\int_{10d} (|L_o|^2+|{\tilde L}_o|^2)\, .
\eea
We now use the decomposition (\ref{decomp-11d}) in the action (\ref{action-11d}), and integrate over the 11th coordinate to get a 10d version of the bulk action. Adding the terms from the boundary action we have
\bea
S_{10d}\sim -\int_{10d} (K_e\, d{\tilde L}_o+{\tilde K}_e\, dL_o)+\frac{\Theta_A}2\int_{10d}(|K_e|^2+|{\tilde K}_e|^2)+\frac{\Theta_B}2\int_{10d} (|L_o|^2+|{\tilde L}_o|^2)\, ,\quad
\eea 
where we have introduced the step functions $\Theta_A$, $\Theta_B$, which are $0,1$ so as to select the correct boundary condition for 10 type 0A or 0B theories. In the context of the cobordism domain wall between the 10d 0A/0B theories, a smoothed out version of the step functions would arise in terms of the tachyon kink. For instance, choosing $\partial^2_X {\cal T}=\frac 2\pi\arctan X^9$, we have 
\bea
\Theta_{A,B}=\frac 12 (1 \pm \partial^2_X {\cal T})\, .
\eea
This provides a dynamical realization of the action introduced (in the type IIA/IIB domain wall context) in \cite{Heckman:2025wqd} to describe the behavior of the RR fields across the domain wall. For completeness we quickly review the argument here.
The resulting equations of motion read
\bea
& \Theta_A * K_e -d{\tilde L}_o=0\quad ,\quad &  \Theta_A * {\tilde K}_e -dL_o=0\, ,\nonumber\\
& \Theta_B * L_o -d{\tilde K}_e=0\quad ,\quad &  \Theta_B * {\tilde L}_o -dK_e=0\, .
\eea
This implies that on the type 0A side, the even-degree RR fields $K_e$, ${\tilde K}_e$ are field strengths of the odd-degree RR potentials $L_o$, ${\tilde L}_o$, whereas in the type 0B side these same odd-degree fields $L_o$, ${\tilde L}_o$ are the field strengths of the even-degree RR potentials $K_e$, ${\tilde K}_e$.

Hence, we recover the description of RR fields in \cite{Heckman:2025wqd} from our 11d perspective. Conversely, our above argument can be regarded as a 11d rewriting of the 10d behavior of RR fields across the wall, supporting our description of the GSO domain wall in terms of supercritical strings. Note also that the $\Z_2$ symmetry for RR fields emphasized in \cite{Heckman:2025wqd}, is reinterpreted in our language as the $\Z_2$ symmetry associated to the 11th dimension.

\section{\texorpdfstring{$(1,1)$}{(1,1)} completion of 2d four-fermion interactions}
\label{app:one-comma-one}

In this appendix we construct the $(1,1)$ supersymmetric completion of the interacting fermionic 2d theories used in section \ref{sec:typeII} in the discussion of the type IIA/IIB domain wall.

We are interested in 2d $(1,1)$ theories whose dynamics reproduces four-fermion interactions. If the four-fermion interactions are considered as fundamental, their supersymmetric completion requires the inclusion of higher-derivative terms (via higher F-terms, see \cite{Beasley:2004ys,Garcia-Etxebarria:2008mpu,Uranga:2008nh} for 4d analogues). A more useful possibility is to consider the four-fermion interactions as effectively mediated by exchange of a massive scalar with Yukawa couplings to the fermions. This version admits a simpler $(1,1)$ supersymmetric completion involving only 2-derivative terms, with the Yukawa couplings encoded in a superpotential. In the following we use this procedure to provide the $(1,1)$ version of the $SO(N)$ Gross-Neveu model; the generalization to other four-fermion theories is similar, and in particular the $Spin(7)$ four-fermion interaction is constructed and used in section \ref{sec:smg}.

We now review the $(1,1)$ version of the 2d $SO(N)$ Gross-Neveu model, with the four-fermion interactions arising from exchange of extra bosons with Yukawa couplings to the fermions (also known as Gross-Neveu-Higgs model).

We have a 2d theory with (1,1) susy and $N$ bosons $\phi^i$ and $N$ Majorana fermions $\psi^i$, which we combine in superfields $\Phi^i$. The superspace lagrangian is (with implicit sums in $i$)
\bea
S=\int d^2x d^2\theta \left[\frac 12 |D\Phi^i|^2+\frac{g^2}{8N}(\Phi^i\Phi^i)^2
\right]\, ,
\label{susy-gn}
\eea
This can be rewritten (with the susy version of the Hubbard–Stratonovich transformation) introducing an extra supermultiplet $\Sigma$, with the action becoming
\bea
S=\int d^2 x d^2\theta \left[\frac 12 |D\Phi^i|^2+\frac 12 \Sigma \Phi^i\Phi^i-\frac{N}{2g^2}\Sigma^2
\right]\, .
\label{action-sigma}
\eea
It is easy to recover (\ref{susy-gn}) by using the equations of motion for $\Sigma$, given by
\bea
\Sigma=\frac{g^2}{2N}\Phi^i\Phi^i\, .
\label{defsigma}
\eea
The action for $\Phi^i$ is quadratic
\bea
S_\Phi=\int d^2x d^2\theta\,\frac 12 \Phi^i(-D^2+\Sigma)\Phi^i\, .
\eea
One can integrate them out to get an effective action for $\Sigma$
\bea
&&e^{-S_{eff}[\Sigma]}=\int {\cal D}\Phi^i e^{-S_\Phi[\Phi^i,\Sigma]}\, ,\\
&&S_{eff}[\Sigma]=\int d^2xd^2\theta\bigg[\, -\frac N{2g^2}\Sigma^2 +\frac N2\, {\rm Str}\,\log\,(-D^2+\Sigma)\,\bigg]\, ,
\eea
where Str is the supertrace over superspace and spinor indices. We take $\Sigma$ to be a slowly varying background superfield and evaluate the
functional trace for a constant $\Sigma$ in superspace. A convenient way to do
this is to differentiate with respect to $\Sigma$:
\begin{equation}
\frac{\delta}{\delta\Sigma}\,{\rm Str}\,\ln(-D^2+\Sigma)
= {\rm Str}\, \frac{1}{-D^2+\Sigma}.
\end{equation}

For constant $\Sigma$, we use the superspace identity (in momentum space)
\begin{equation}
(-D^2+\Sigma)(D^2+\Sigma)=p^2+\Sigma^2,
\end{equation}
which implies
\begin{equation}
\frac{1}{-D^2+\Sigma}=\frac{D^2+\Sigma}{p^2+\Sigma^2}.
\end{equation}

In the superspace trace, only the $D^2$ term acting on the Grassmann delta
function contributes, since the $\Sigma\,\delta^2(\theta-\theta')$ term
vanishes at coincidence. Therefore,
\begin{equation}
{\rm Str}\,\frac{1}{-D^2+\Sigma}
= \int d^2x\,d^2\theta \int_{|p|<\Lambda}\frac{d^2p}{(2\pi)^2}\,
\frac{1}{p^2+\Sigma^2}.
\end{equation}

With a hard momentum cutoff $\Lambda$ in two dimensions,
\begin{equation}
\int_{|p|<\Lambda}\frac{d^2p}{(2\pi)^2}\frac{1}{p^2+\Sigma^2}
= \frac{1}{4\pi}\ln\frac{\Lambda^2+\Sigma^2}{\Sigma^2}.
\end{equation}

Integrating back with respect to $\Sigma$ and dropping $\Sigma$-independent
terms, we obtain
\begin{equation}
{\rm Str}\,\ln(-D^2+\Sigma)
= \int d^2x\,d^2\theta\left[
\frac{\Lambda}{2\pi}\arctan\!\left(\frac{\Sigma}{\Lambda}\right)
+\frac{\Sigma}{4\pi}\ln\frac{\Lambda^2+\Sigma^2}{\Sigma^2}
\right].
\end{equation}

The effective action then looks like

\begin{equation}
S_{\text{eff}} \;=\; \int d^2x\, d^2\theta\; U(\Sigma),
\end{equation}
\begin{equation}
U(\Sigma)
= -\,\frac{N}{2g^2}\,\Sigma^2
+ \frac{N}{2}\left[
\frac{\Lambda}{2\pi}\,\arctan\!\left(\frac{\Sigma}{\Lambda}\right)
+ \frac{\Sigma}{4\pi}\,\ln\!\frac{\Lambda^2+\Sigma^2}{\Sigma^2}
\right].
\end{equation}
The path integral over $\Sigma$ then forces the condition $U'(\Sigma)=0$. In terms of a vev $\langle\Sigma\rangle=\sigma$, this leads to the gap equation

\begin{equation}
\frac{\sigma}{g^2}
=\frac{1}{8\pi}\ln\!\left(1+\frac{\Lambda^2}{\sigma^2}\right).
\end{equation}
The solution of this equation (call it $\sigma_*$) gives a mass gap. From (\ref{defsigma}) and taking expectation values,
\begin{equation}
\langle\Sigma\rangle=\frac{g^2}{2N}\,\langle\Phi^i\Phi^i\rangle.\label{stuff}
\end{equation}

Now expanding the real $(1,1)$ superfields in components
\begin{equation}
\Phi^i
=\phi^i+\theta^+\psi_+^i+\theta^-\psi_-^i+\theta^+\theta^-F^i,
\qquad
\Sigma
=\sigma+\theta^+\chi_+ +\theta^-\chi_-+\theta^+\theta^-F_\Sigma\, ,
\end{equation}
and plugging back in \eq{stuff}, we get
\begin{equation}
\langle\psi_+^i\psi_-^i\rangle
=2\,\langle\phi^iF^i\rangle=-2\sigma_*\,\langle\phi^i\phi^i\rangle=-\frac{4N}{g^2}\sigma_*^2\, ,
\end{equation}
so a fermion condensate develops, just like in the non-supersymmetric Gross-Neveu model. $\sigma_*$ also controls the dynamical fermion mass.

\bibliographystyle{utphys}
\bibliography{refs.bib}

\end{document}